\documentclass[reprint,aps,prl,nofootinbib]{revtex4-1}

\usepackage{graphicx}
\usepackage{caption}
\usepackage{subcaption}
\usepackage{float}
\usepackage{xcolor}
\usepackage{amssymb,amsmath,amsthm}
\usepackage{algorithm}
\usepackage{algpseudocode}
\usepackage{url}
\usepackage[colorlinks,
            pdffitwindow=false,
            plainpages=false,
            pdfpagelabels=true,
            pdfpagemode=UseOutlines,
            pdfpagelayout=SinglePage,
            bookmarks=false,
            colorlinks=true,
            hyperfootnotes=false,
            linkcolor=blue,
            urlcolor=blue!50!black,
            citecolor=green!50!black]
{hyperref}
\usepackage{comment}
\usepackage{bm}

\newcommand{\R}{\mathbb{R}}
\newcommand{\N}{\mathbb{N}}
\newcommand{\E}{\mathbb{E}}

\newcommand{\numA}{K}  % number of anchor points
\newcommand{\anchor}{k}  % index of anchor point
\newcommand{\numS}{S}  % number of samples per anchor point
\newcommand{\sample}{s}  % index of sample
\newcommand{\cv}{\varphi}  % collective variable
\newcommand{\ecv}{\bar{\varphi}}  % extended CV
\newcommand{\cvmat}{\Phi}  % matrix of CV evaluations
\newcommand{\ecvmat}{\bar{\Phi}}  % matrix of extended CV evaluations
\renewcommand{\vec}[1]{\bm{#1}}  % vectors bold

\newcommand{\kernel}{\kappa}
\newcommand{\kernelmat}{M}

\begin{document}
\title{Learning Interpretable Collective Variables for Spreading Processes on Networks}

\author{Marvin Lücke}
\affiliation{Zuse Institute Berlin}

\author{Stefanie Winkelmann}
\affiliation{Zuse Institute Berlin}

\author{Jobst Heitzig}
\affiliation{FutureLab on Game Theory and Networks of Interacting Agents, Potsdam Institute for Climate Impact Research}
\affiliation{Zuse Institute Berlin} % habe ich neuerdings auch

\author{Nora Molkenthin}
\affiliation{Complexity Science Department, Potsdam Institute for Climate Impact Research}

\author{Péter Koltai}
\affiliation{Department of Mathematics, University of Bayreuth}

\begin{abstract}
Collective variables (CVs) are low-dimensional projections of high-dimensional system states. They are used to gain insights into complex emergent dynamical behaviors of processes on networks.
The relation between CVs and network measures is not well understood and its derivation typically requires detailed knowledge of both the dynamical system and the network topology.
% In processes on networks, collective variables are often observed to be related to network measures. However, the nature of this relation is not well understood and its derivation requires detailed knowledge of both the dynamical system and the network topology.
In this work, we present a data-driven method for algorithmically learning and understanding CVs for binary-state spreading processes on networks of arbitrary topology.
We demonstrate our method using four example networks:
the stochastic block model, a ring-shaped graph, a random regular graph, and a scale-free network generated by the Albert--Barabási model. Our results deliver evidence for the existence of low-dimensional CVs even in cases that are not yet understood theoretically.
% Our method extends the recent transition manifold approach to produce interpretable CVs that are linked to understandable and generalizable network features.
\end{abstract}

\maketitle

\paragraph{Introduction.}
Networks of interacting agents are widely used to model socio-dynamical phenomena \cite{Castellano2009} such as the spreading of a disease \cite{Kiss2017, PastorSatorras2015, Mauras2021} or the diffusion of a (political) opinion within a society~\cite{Banisch2019, das2014}.
In such networks, nodes represent individual agents, and edges represent some form of social interaction.
Each node has a state that evolves over time depending on the states of neighboring nodes.
Often, stochastic effects are included to account for uncertainty in the dynamics and for the unpredictability of agents.
These types of spreading processes are at the core of numerous open problems in a wide range of disciplines, such as understanding social collective behavior \cite{BakColeman2021}, assessing systemic risk in financial systems \cite{Caccioli2017}, or controlling modern power grids~\cite{Witthaut2022}.

%This paper investigates a variant of the so-called ``voter model'', the \emph{continuous-time noisy voter model} (CNVM), in which agents switch stochastically between discrete states based on their neighborhoods.
%Although the interaction rules are quite simple, the emergent macroscopic behavior of the system remains complex and difficult to analyze.
%Depending on model parameters and network structure, a rich variety of behaviors can be observed~\cite{Carro2016, Castellano2009, Granovsky1995, Presutti1983}.

One approach to elevate our understanding of these models is to seek a low-dimensional representation of the system that captures the fundamental dynamics on timescales of interest.
The projection into this low-dimensional space is called a \emph{collective variable} (CV), and the projected dynamics is called the \emph{effective dynamics}.
Good CVs retain the essential information about the system's behavior, reducing the dimensionality and enabling a more efficient analysis and prediction.

For discrete-state spreading processes on certain simple networks, e.g., complete graphs and dense Erd\H{o}s--Rényi random graphs, it is known that the shares of nodes in each state constitute a good CV, and its evolution is given by an ordinary differential equation in the mean-field limit~\cite{Luecke2022}.
There are many other results, valid for a varying range of networks and models, that describe the evolution of the shares of states in terms of (partial) differential equations in the mean-field or hydrodynamic limit~\cite{Presutti1983, keliger2022, Durrett2016, Fan2021}.
Another popular choice of CVs are the counts of certain network motifs, e.g., the number of links between nodes of different states (typically called pair-approximation~\cite{keeling1997correlation,vazquez2008analytical,pugliese2009heterogeneous}), and moment-closure methods can be used to approximate their evolution~\cite{do2009contact, taylor2012markovian, Moretti2012, Kuehn2016, Peralta2018}.
% \pk{Add more MF, PA cites here?}. 
For the special case of binary-state dynamics, standard mean-field or pair-approximation theories have been complemented by higher-order master equations which improve accuracy by introducing fractions of neighbors of one or the other state, resolved with respect to the degrees of the nodes~\cite{eames2002modeling,marceau2010adaptive,gleeson2011high,lindquist2011effective,gleeson2013binary}. 
%For the gain of more accuracy, Approximate Master Equations~\cite{gleeson2011high,gleeson2013binary} introduce fractions of neighbors of one or the other state, resolved with respect to the degrees of the nodes.
However, there is no all-encompassing theory relating any network topology and any process occurring on it to resulting~CVs.
Hence, a constructive computational approach -- like the one we will present -- can elucidate cases that theoretical results do not yet cover.
Moreover, the above-mentioned techniques \emph{postulate} a candidate CV based on (physical) insights about the system, whereas our procedure does not require such intuition.
%An important feature of our procedure is that no such intuition is required. Nonetheless, as we will demonstrate, it may be beneficial when interpreting the constructed~CV.

\begin{figure*}[htb]
\includegraphics[width=\textwidth]{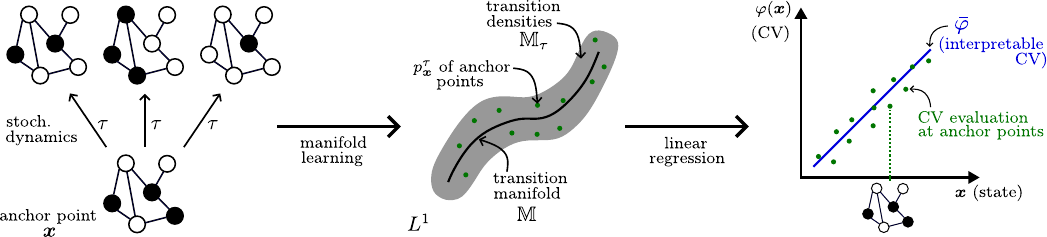}
\caption{Illustration of our method. Left: The random process is described by its distribution, sampled through $S$ samples per initial network state (anchor). Middle: These distributions are used to learn a low-dimensional parametrization of the transition manifold. Right: A regression step allows for interpretability of the learned~CV.}
\label{fig:method}
\end{figure*}

In this work, we extend the \emph{transition manifold approach} \cite{Bittracher2017, Bittracher2020} to learn CVs based on simulation data and to systematically find the relationship of the learned CVs to topological features of the network (see Fig.~\ref{fig:method}).
The transition manifold approach assumes and exploits that the transition density functions of the system accumulate around a low-dimensional manifold, from which a CV can be inferred.
The approach is designed such that most information of the density propagation of the process is retained under projection onto the CV~\cite{Bittracher2017}.
To make it applicable to binary-state spreading processes on networks, we develop a technique for evenly sampling the state space and we add a linear regression step to produce interpretable CVs.
% Furthermore, we combine the transition manifold approach with a linear regression step to produce an interpretable CV for binary-state spreading processes, from which we can evaluate the role and importance of each node in the network.

While goals similar to ours have been pursued in the literature before, to our best knowledge, this is the first work to learn CVs from data for networks of interacting agents. 
A crucial difference to previous works which learn reduced (also called surrogate or emergent) spaces~\cite{coifman2008diffusion,rohrdanz2011determination,berry2015nonparametric,KoWe20} from data -- even in the agent-based context~\cite{kemeth2022learning} -- is that they perform the reduction primarily based on the information of a state, and not on its dynamics. 
Once a good CV has been learned with our method, a surrogate dynamical model for its evolution could be determined and analyzed by tools also utilized in these references, in particular by linear transition operators associated with the dynamics~\cite{lusch2018deep,mardt2018vampnets,froyland2021spectral}. 
% This is deferred to future studies.
%The data-driven nature of our method could even allow its usage if no model and network but only adequate observations are given. 
Other approaches for data-based analysis and reduced modelling of interacting agent systems postulate CVs instead of learning them~\cite{lu2019nonparametric,maggioni2019data,Niemann2021, WuKoSch21,Helfmann2021}.

Recently, deep learning techniques have become popular for finding low-dimensional variables and surrogate dynamical models~\cite{brandt2018machine,wehmeyer2018time,lusch2018deep,mardt2018vampnets,raissi2018multistep,chen2019nonlinear,champion2019data,otto2019linearly}.
Artificial neural networks can represent coordinates from a large general class, but the dynamical conditions necessary for them to perform well remain implicit in these methods.
Our approach, however, relies on explicit dynamical assumptions that are validated during the data-driven computation.

\paragraph{The transition manifold.} 
Consider a fixed network of $N$ nodes, on which each node $i \in \{1,\dots,N\}$ has a discrete state $x_i \in \mathbb{S}$.
The state space of the process is thus $\mathbb{X} := \mathbb{S}^N$, and its elements are system states $\vec{x} = (x_1,\dots,x_N)$.
Given a system state $\vec{x}\in \mathbb{X}$ and time $\tau \geq 0$, the \emph{transition density function} $p_{\vec{x}}^\tau \in L^1(\mathbb{X}) =: L^1$ is defined such that $p_{\vec{x}}^\tau(\vec{y})$ is the probability that the system is in state $\vec{y}$ at time $\tau$ after having started in state $\vec{x}$ at time~$0$. (The term \textit{density} comes from the original theory for continuous state spaces. We use $L^1$ to emphasize that these are vectors with entries summing to~1.)
The \textit{transition manifold approach} \cite{Bittracher2017, Bittracher2020} exploits the observation that for certain systems and an appropriate choice of $\tau$, the set
\begin{equation}
    \mathbb{M}_\tau := \{p_{\vec{x}}^\tau \mid \vec{x} \in \mathbb{X}\} \subset L^1
\end{equation}
is close to a $d$-dimensional submanifold $\mathbb{M} \subset L^1$ called the \emph{transition manifold}.
The lag-time $\tau$ needs to be longer than the ``relaxation time'' towards $\mathbb{M}$ and shorter than the time it takes to eventually converge to a stationary distribution (see supplemental material \cite[section S.2]{supplemental} for details).

As a consequence, one can show that there exists a $d$-dimensional \textit{collective variable} $\cv: \mathbb{X} \to \R^d$, such that for all $\vec{x} \in \mathbb{X}$
\begin{align}
    p_{\vec{x}}^\tau \approx \tilde{p}_{\cv(\vec{x})}^\tau, \label{eq:tilde_p}
\end{align}
for some function $\tilde{p}_{(\cdot)}^\tau$. Hence, the essential information needed to characterize the dynamics is captured by the collective variable~$\cv$.
A coordinate function $\cv$ satisfying this is for instance a ``parametrization'' of the ma\-ni\-fold $\mathbb{M}$, in the sense that the assignment $\mathbb{M} \to \mathbb{X} \to \mathbb{R}^d,\   p_{\vec{x}}^\tau \mapsto \vec{x} \mapsto \cv(\vec{x})$ is one-to-one, cf.~\cite{Bittracher2017,Bittracher2020}.
Typically, the dimension $d$ of the reduced state is significantly smaller than the dimension of the original state. 

From now on, we consider binary-state dynamics, i.e., $\mathbb{S}=\{0,1\}$.
While this already covers a wide range of interesting dynamics, we expect that most of the following can be extended to an arbitrary number of states, at the cost of additional technicalities.
% (often referred to as ``susceptible'' and ``infected'').
For binary-state dynamics on a complete network, to give an example, a good collective variable is often given by the share of nodes in state 1, i.e., $d = 1$ and $\cv(\vec{x}) = \sum_i x_i / N$ \cite{Luecke2022}.
This is related to the quantity called magnetization in the similar Ising model \cite{Cipra1987}.

%For example, on a complete graph the collective variable of a voter model \sw{voter model does not appear at all anymore... Here, we should replace it by binary state dynamics, but later we have to introduce it for our examples} is given by the share of nodes in state 1, i.e., $d = 1$ and $\cv(\vec{x}) = \sum_i x_i / N$ \cite{Luecke2022}. In the context of the similar Ising model \cite{Cipra1987}, this quantity is also called \textit{magnetization}.

\paragraph{Learning interpretable CVs.} 
We propose the following method, which consists of three steps (Fig.~\ref{fig:method}).
% \sw{I find the following intro/summary unnecessary because now the steps are anyway very shortly explained afterwards.}
% First, we choose a diverse set of states, which are called the \textit{anchor points}.
% Then, we simulate from each anchor point a fixed number of short trajectories and approximate the transition manifold based on that data.
% Finally, we employ linear regression to learn an interpretable CV by fitting the transition manifold data from the previous step .

% \begin{figure*}[t]
% \includegraphics[width=\textwidth]{method.pdf}
% \caption{The method.The method.The method.The method.The method.The method.}
% \label{fig:method}
% \end{figure*}

\emph{Step 1.}
We choose a diverse set of dynamically relevant \emph{anchor points} $\vec{x}^1, \dots, \vec{x}^\numA \in \mathbb{X}$ in which the CV is going to be computed in the first instance. Diversity of the points is crucial in the sense that their respective transition densities cover $\mathbb{M}_\tau$ sufficiently well.
Otherwise, the parametrization learned in the next step would yield a result biased by the insufficient coverage, that is potentially not a CV for the entire process.
A precise quantification of sufficient coverage cannot be stated in generality as it depends on the system at hand and on the desired quality of the CVs. We employ an algorithm that prioritizes sampling anchor points containing communities of nodes with the same state, as this strategy produced the best results for the spreading processes we examined, see supplemental material~\cite[section S.2]{supplemental} for details.
% The algorithm we use to sample these diverse anchor points is described in detail in the supplemental material~\cite[section S.2]{supplemental}.

\emph{Step 2.}
Next, we approximate a ``parametrization'' $\cv$ of the transition manifold $\mathbb{M}$ from simulation data.
% To this end, most efficient methods require the computation of local distances.
% A particularly advantageous distance between densities $p^{\tau}_{\vec{x}}$ and $p^{\tau}_{\vec{y}}$ is the \emph{maximum mean discrepancy} (MMD), as it can be efficiently approximated using samples of the densities \cite{Bittracher2020, Muandet2017}.
For each anchor point $\vec{x}^\anchor$ we conduct $\numS \in \mathbb{N}$ simulations of the process of length $\tau$, yielding $\numS$ samples for each transition density $p^{\tau}_{\vec{x}^\anchor}$.
% Using this data we approximate the \textit{distance matrix} $\Delta \in \R^{\numA\times \numA}$ that contains the pairwise MMDs between the densities associated to the anchor points.
Using this data, we obtain evaluations of the collective variable $\cv$ at the anchor points $\vec{x}^1,\dots,\vec{x}^\numA$ by applying a manifold learning technique. Here, the dimension $d$ of the CV is also an output.
% algorithm to $\Delta$. We choose the diffusion maps method \cite{Coifman2006} for this purpose.
See supplemental material \cite[section S.2]{supplemental} for further details on the approximation of the transition manifold.

The necessary number $\numA$ of anchor points and $\numS$ of simulations per anchor point depend on the size and complexity of the network. For the examples below ($N < 1000$), we found $\numA \approx 1000$ and $\numS \approx 100$ adequate.
This very small number of anchor points compared to the number $2^N$ of all possible states is sufficient due to the targeted sampling method we employ (cf. step 1).
We observed a substantial robustness of the results in varying the method's hyperparameters, but an entirely automatic procedure for their selection has yet to be designed.
% Furthermore, we propose to choose a lag time $\tau$ so that nodes are expected to experience at least one state transition, i.e., $\tau$ of order $(r_{m,n} + \tilde{r}_{m,n})^{-1}$.
% from the diffusion maps algorithm~\cite{CSSS08}, or can be inferred by inspecting the network structure and plots of the transition manifold.
% Moreover, it is computationally cheap to test different dimensions~$d$ after the distance matrix $\Delta$ has been computed.
The output of this second step of the method is evaluations of the $d$-dimensional CV $\cv$ at the anchor points, $\cv(\vec{x}^1), \dots, \cv(\vec{x}^\numA) \in \R^d$.

\emph{Step 3.} 
The third step of the method aims at determining the meaning of the CV and finding a reasonable map $\ecv: \{0,1\}^N \to \R^d$ that extends it to states $\vec{x}$ outside of the original data set.
Motivated by the fact that, for binary-state dynamics, the share of nodes in state $1$ in (parts of) the network is known to be a good CV in specific cases~\cite{Luecke2022}, we propose maps $\ecv$ of the form
\begin{align}
    \ecv(\vec{x}) = \begin{pmatrix}
\ecv_1(\vec{x})\\
\vdots \\
\ecv_d(\vec{x})
\end{pmatrix}, \quad \ecv_j(\vec{x}) = \sum_{i=1}^N \Lambda_{j, i}\; x_i,
\label{eq:zeta}
\end{align}
where $\Lambda \in \R^{d\times N}$ is a parameter matrix.
For example, choosing $d=1$ and $\Lambda = (1,\dots,1)$ yields a map describing the total count of state~$1$.
In the different context of coupled ODEs, a CV similar to \eqref{eq:zeta} was examined in Refs.~\cite{Gao2016, Laurence2019}.

We find optimal parameters $\Lambda$ by employing linear regression to fit $\ecv$ to the computed CV values in the anchors, $\cv(\vec{x}^1), \dots, \cv(\vec{x}^\numA)$, from step 2.
To prevent overfitting we use the \emph{graph total variation} regularizer, which penalizes variation of $\Lambda$ along edges (see supplemental material \cite[section S.3]{supplemental}).
The reason for this choice is that each node in densely interconnected clusters is expected to contribute similarly to the CV. Even for networks containing hubs we can maintain this regularizer by bootstrapping our result to modify the ansatz functions, cf.\ Example~4 below.
There we use linear basis functions for a first regression step, observe strong correlation in $\Lambda$ to the network's degrees, and modify the ansatz functions accordingly.
If suggested by physical intuition, one can also entirely deviate from the linear ansatz functions we propose. % from the start, although we will see that~\eqref{eq:zeta} is a good default choice.

In the following examples binary-state spreading processes are studied, in which each node experiences memoryless random evolution in continuous time, with transition rates determined by a constant exploration rate (noise) and an ``influence'' rate based on the states in the node's neighborhood, see supplemental material~\cite[section S.1]{supplemental} for details.
%The following examples deal with binary-state spreading processes on different kinds of networks. 
% In the processes studied, each node $i$ experiences memoryless random evolution in continuous time, with transition rates determined by a constant exploration rate (noise) and an ``influence'' rate based on the states in the node's neighborhood, see supplemental material~\cite[section S.1]{supplemental} for details.
The investigations below refer to the \textit{noisy voter model}~\cite{Granovsky1995,Carro2016}, yet the supplemental material also includes findings related to alternative dynamics.
%In the processes considered, each node $i$ undergoes a memoryless random evolution in continuous time, where the transition rates governing the probabilities to switch to a different state $x_i$ within some time are composed from a constant exploration rate and an ``influence'' rate that is a function of the shares of the states in the node's neighborhood. 
%For details, see~\cite[section S.1]{supplemental}. \sw{Still mention the term voter model?}
%\pk{We could say that it is the basis for our examples here and the SM includes results for other dynamics.}

\begin{figure}[b]
     \centering
     \includegraphics[width=\linewidth]{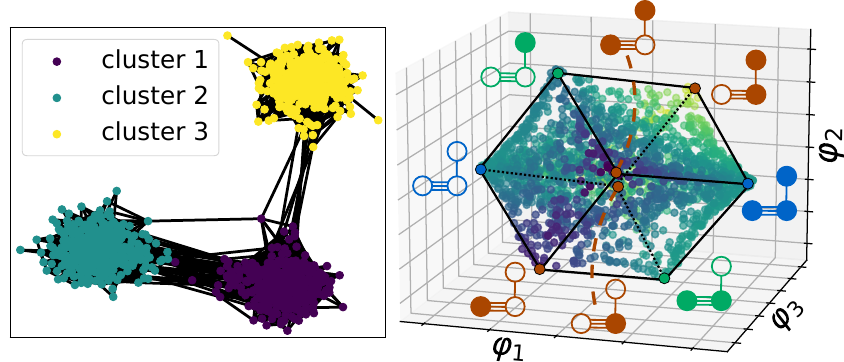}
    \caption{For the stochastic block model network (left), the transition manifold is a 3-dimensional cuboid (right). The vertices of the cuboid correspond to extreme states $\vec{x}$ where for each cluster either all (filled circle) or no nodes (empty circle) have state~$1$.}
    \label{fig:sbm}
\end{figure}

\begin{figure}
    \centering
    \includegraphics[width=\linewidth]{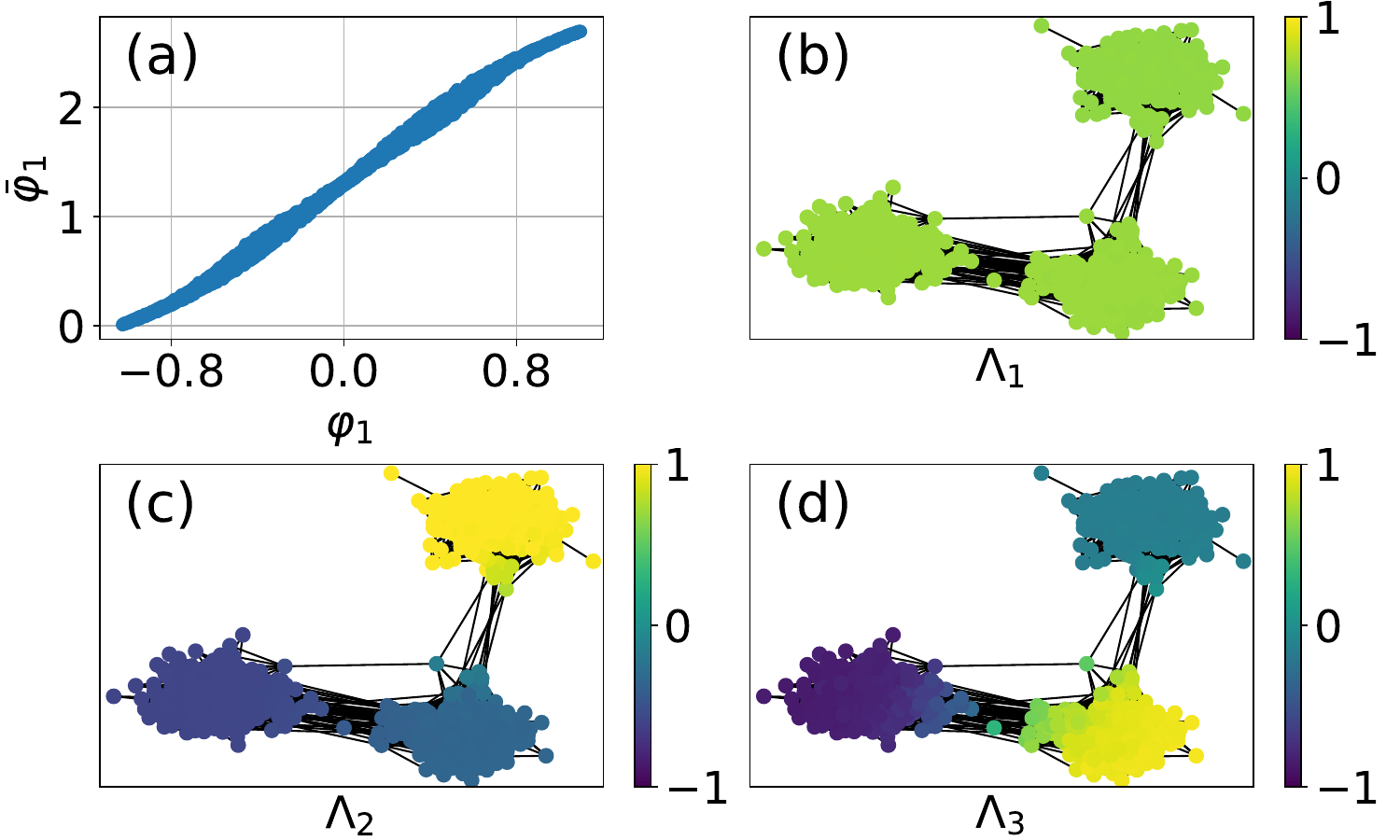}
    \caption{Optimal $\Lambda$ from~\eqref{eq:zeta} for Example~1. (a): Data $\cv_{1}(\vec{x}^k)$ versus optimal fit~$\ecv_1$. (b)-(d): Optimal $\Lambda$ entries for the respective coordinates plotted as color values on the network.}
    \label{fig:sbm_cv}
\end{figure}

\paragraph{Example 1: Stochastic block model.}
We examine a network of $N=900$ nodes that is constructed using the \textit{stochastic block model}.
The network consists of three clusters such that cluster 1 and 2 are densely connected, cluster 1 and 3 are connected only sparsely, and cluster 2 and 3 are not connected at all, cf.\ Fig.~\ref{fig:sbm}.
We expect the optimal CV to be $d=3$-dimensional and contain the counts of 1's in each cluster. This CV is exact in the sense that for $N\to \infty$ it satisfies a mean-field equation~\cite{Luecke2022}.
Applying our method and plotting the resulting CV point cloud $\{\cv(\vec{x}^1),\dots, \cv(\vec{x}^\numA)\} \subset \R^3$ yields an approximately cuboid shaped transition manifold.
We found that the vertices of this cuboid correspond to extreme states $\vec{x}$ in which for each cluster either all or no nodes have state $1$, cf.\ Fig.~\ref{fig:sbm}.
To discover the meaning of the three coordinates $\cv_1, \cv_2, \cv_3$, we calculate the optimal fit according to the regression problem proposed in step 3 of the method, which yields a collective variable $\ecv(\vec{x}) = \Lambda \vec{x}$ with optimal parameters $\Lambda$ shown in Fig.~\ref{fig:sbm_cv}.
The entries of the first row $\Lambda_{1,:}$ are approximately equal and thus $\ecv_1$ describes the count of 1's in the whole network.
The optimal $\Lambda_{2,:}$ is positive and constant within cluster 3 and negative and constant within clusters 1 and~2.
Thus, $\ecv_2$, calculates how the 1's are distributed between clusters $\{1, 2\}$ and~$\{3\}$.
% (For example, a large positive value of $\ecv_2$ indicates that there are many 1's in cluster 3 and few in clusters 1 and~2).
Finally, $\Lambda_{3,:}$ is positive in cluster 1, negative in cluster 2, and approximately 0 in cluster 3, which implies that $\ecv_3$ measures how the 1's are distributed between clusters $1$ and $2$, regardless of the number of 1's in cluster~3.
Hence, the learned CV $\ecv$ includes exactly the information that was predicted by theory~\cite{Luecke2022} for large~$N$, i.e., the counts of 1's for each cluster, but the coordinates are ordered by dynamical prevalence.
% However, the transition manifold method learned a transformation of this information such that the coordinates are ordered by dynamical prevalence.
(For instance, coordinate 3 is the least prevalent because information flows quickly between the two densely connected clusters 1 and~2.)

An interesting question would be to consider the change of the CV and especially its dimension with increasing edge density between the clusters. In particular, do structural transitions in the CV coincide with the so-called \emph{detectability threshold} of the stochastic block model~\cite{reichardt2008detectable,decelle2011inference,banks2016information}, where the edge statistics become indistinguishable from an Erd{\H o}s--R\'enyi random graph model? This will be addressed in future work.
% the first and most important coordinate describes the global count, the second coordinate the distribution of 1's between the two densely connected clusters $\{1, 2\}$ and the sparsely connected cluster 3, and the third coordinate the distribution between clusters 1 and 2.
% The third coordinate is the least important because information flows quickly between the two densely connected clusters 1 and 2, and hence the dynamics equilibrates quickly. \sw{Here, one could shorten a little bit because of repetition?}

\paragraph{Example 2: Ring-shaped network.}
We apply our method to a ring-shaped network of $N=50$ nodes.
Examining the point cloud $\{\cv(\vec{x}^1),\dots, \cv(\vec{x}^\numA)\}$ for different choices of $d$, we can not identify a low-dimensional transition manifold as increasing $d$ keeps adding valuable information.
To keep the CV dimension reasonably small, we choose $d=5$.
(This yields CVs of reasonable quality, cf. supplemental material
\cite[section S.5]{[{See Supplemental Material }][{.}]supplemental}.)

\begin{figure}
     \centering
     \includegraphics[width=.85\linewidth]{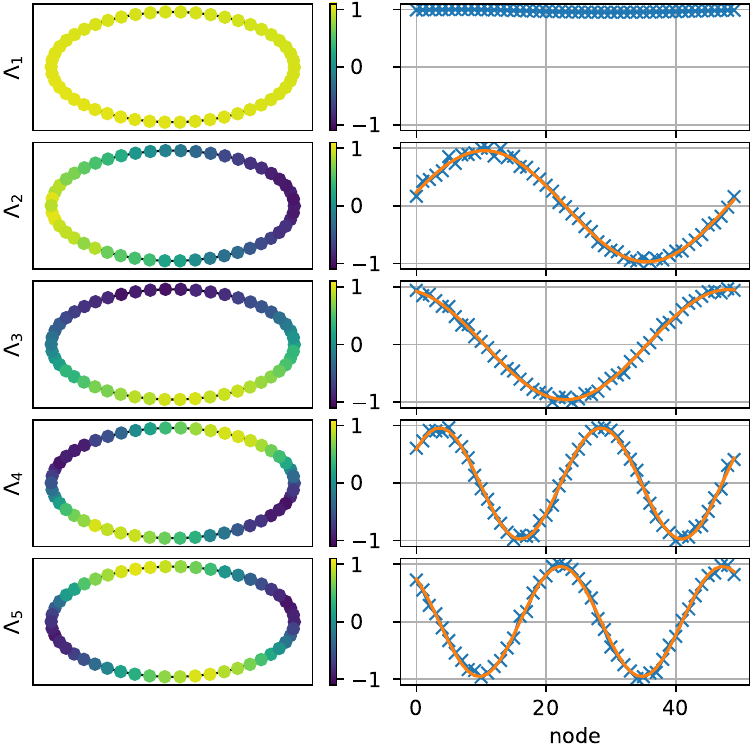}
    \caption{Left: optimal $\Lambda_{i,:}$ plotted as color values on the ring-shaped network. Right: $\Lambda_{i,:}$ (blue crosses) and a sine fit (orange line). The collective variables $\cv_i$ represent the real Fourier coefficients of the distribution of 1's on the ring, since $\cv_i(\vec{x}) \approx \Lambda_{i,:} \, \vec{x}$ with the $\Lambda_{i,:}$ being sines and cosines of increasing frequencies.}
    \label{fig:ring}
\end{figure}

Solving the regression problem in step 3 yields a $\Lambda$ that is constant in the first coordinate, i.e., the most important information is again the total count of 1's, see Fig.~\ref{fig:ring}.
The subsequent $\Lambda_{j,i}$ are pairs of sine and cosine functions of the node index $i$, starting with one oscillation for coordinates $j=2,3$ and then doubling the frequency for coordinates $j=4,5$.
Hence, the collective variable $\ecv$ measures the distribution of 1's on the ring, with increasing precision as we let $d$ increase.
This structure mimics Fourier coefficients, which suggests that (in the limit of infinitely many nodes) the optimal collective variable measures the position-dependent concentration of 1's as a density function on the ring.
This result agrees well with other works considering ring-shaped or lattice networks, e.g.~\cite{Durrett2016, Fan2021, Presutti1983}, which find that the concentration of 1's is governed by a diffusive PDE in the hydrodynamic limit. The CV of the system thus being a function on the ring, any finite-dimensional approximation has a truncation error. However, orthogonal trigonometric polynomials are a natural (and in an $L^2$-sense optimal) choice, found by our method.

\paragraph{Example 3: Random 3-regular network.}
One challenge in reduced modelling of spreading processes on random regular graphs is that edges are correlated. 
If the degree grows indefinitely with the network size, it was shown for the voter model that the share of state 1 is an asymptotically perfect CV~\cite{Luecke2022}. In the same study it was observed numerically that for small degrees this CV still seems to support an effective dynamics, which deviates from the one obtained by mean-field approximation. Our method applied to a random 3-regular network validates the observation by reproducing this~CV; see the supplemental material \cite[section S.4]{supplemental}.

\begin{figure}
    \centering
    \includegraphics[width=\linewidth]{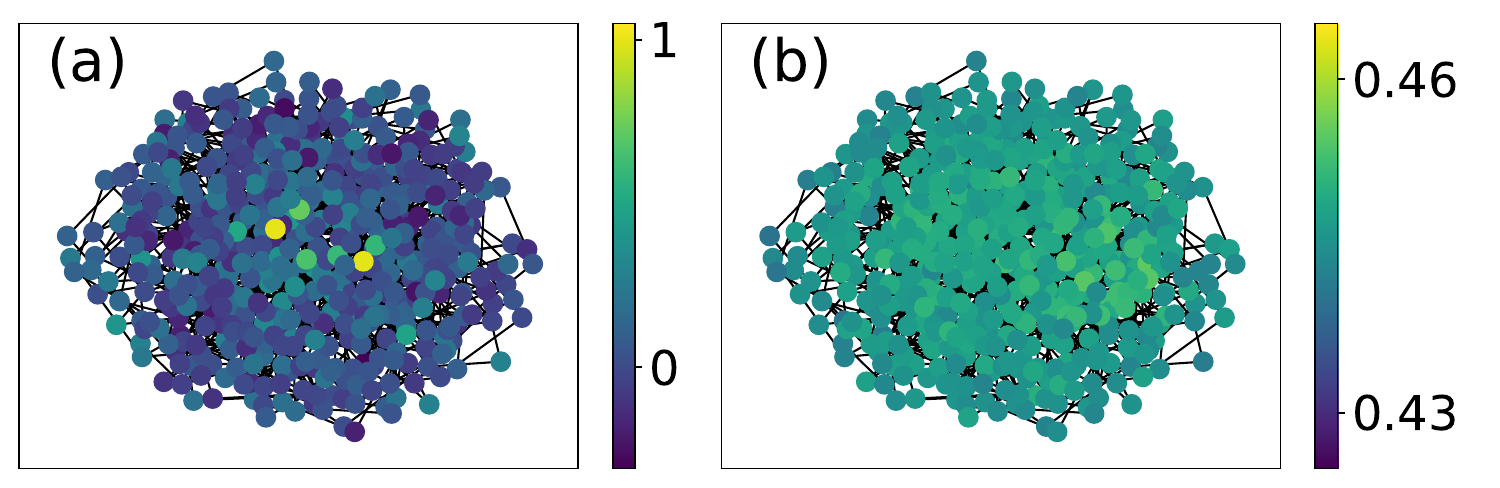}
    \caption{(a) For the Albert--Barabási network, the optimal $\Lambda$ as in~\eqref{eq:zeta} assigns a large weight to nodes with high degree. (b) After pre-weighting with node degree, cf.~\eqref{eq:zeta_weight}, the optimal $\Lambda$ is constant. Hence, the collective variable describes the degree-weighted count of~1's.}
    \label{fig:ba}
\end{figure}

\paragraph{Example 4: Albert--Barabási network.}
Finally, we apply our method to a network generated by the Albert--Barabási model \cite{Barabasi2002}.
In the preferential attachment algorithm each new node is connected to $m=2$ existing nodes, that are randomly picked with probability proportional to their degree. This procedure yields (asymptotically) a scale-free network.
Applying our method results in a point cloud $\{\cv(\vec{x}^1),\dots, \cv(\vec{x}^\numA)\}$ that indicates a $d=1$-dimensional transition manifold, see~\cite[Fig.~S.3]{supplemental}.
The optimal $\Lambda\in \R^N$ according to the linear regression problem in step 3 assigns a large positive weight to nodes of high degree, whereas nodes with small degree have small or even negative weight, cf.\ Fig.~\ref{fig:ba}. This conflicts with our choice of graph total variation regularizer that favors solutions for which $\Lambda$ is equal for neighboring nodes.
We tackle this issue by applying a pre-weighting of each node $i$ with its degree $d_i$:
\begin{align}
    \ecv(\vec{x}) = \sum_{i=1}^N \Lambda_{i}\; d_i\; x_i.
    \label{eq:zeta_weight}
\end{align}
% where $d_i$ denotes the degree of node~$i$.
% Transferring this pre-weighting to the linear regression \eqref{eq:LASSO} yields
% \begin{align}
%     \Lambda = \underset{\vec{\lambda} \in \R^N}{\text{argmin}}\ \lVert E X^T D \vec{\lambda} - E \cvmat \rVert + \alpha \cdot \text{TV}(\vec{\lambda}),
%     \label{eq:LASSO_weight}
% \end{align}
% where $D = \text{diag}(d_1,\dots,d_N) \in \R^{N\times N}$.
The optimal $\Lambda$ for \eqref{eq:zeta_weight} becomes approximately constant, and hence the CV measures the degree-weighted count of state $1$ in the system, cf.\ Fig.~\ref{fig:ba}.
Multiple experiments for varying parameters confirmed this result, provided the preferential attachment parameter is chosen~$m \geq 2$.
(For $m = 1$ the resulting networks exhibit a significantly larger diameter~\cite{Bollobas2004}. As a consequence, the degree-weighted count does not seem to sufficiently characterize the dynamics.)
% We carried out multiple experiments for different realizations of the random networks for varying parameters, and all indicated that the degree-weighted count of state $1$ is generally the best collective variable for Albert--Barabási networks, provided the preferential attachment parameter is chosen $m \geq 2$. For $m = 1$ the resulting networks are trees and exhibit a significantly larger diameter (asymptotically $\log N$ instead of $\log N/\log \log N$ \cite{Bollobas2004}), so that higher dimensional collective variables are needed that measure the concentration of 1's in different parts of the tree.
We are not aware of any theoretical work showing that the degree-weighted count is a good CV for (binary-state or other) spreading processes on Albert--Barabási networks, although refs.~\cite{Suchecki2005,bianconi2002mean} hint at the significance of this observable.

\paragraph{Validation.}
A numerical validation of the CVs learned in the above examples is presented in the supplemental material \cite[section S.5]{supplemental}.

\paragraph{Conclusion.}
We propose a method to learn interpretable CVs for spreading processes on networks without the need for prior expert knowledge about the network topology or the dynamical process.
This method consists of the following steps: First we sample anchor (network) states, from which we start many short simulations. Then we approximate the transition manifold and extend the learned CVs to unseen data using (total-variation-regularized) linear regression.
The CVs are interpretable since the inferred parameters indicate the function and significance of features of the network structure.
We have demonstrated this method for four different network topologies and two types of spreading dynamics (see supplemental material \cite[section S.4]{supplemental}) and have thus shown its flexibility and usefulness. Although out of scope for the current manuscript, we expect that the method can be generalized to processes with more than two discrete states as well as inhomogeneous agent dynamics.
%In future work our method could be applied to different dynamics, e.g., other variants of the voter model or majority rule models, and the impact on the CVs could be investigated.
%It would be interesting to extend the method to models of more than two discrete states, or even systems with a continuous state space.
%Moreover, the presented method to learn collective variables could be extended to also infer their dynamical evolution from data, thus learning a reduced surrogate model of the underlying dynamics.

\begin{acknowledgments}
This work has been supported by Deutsche Forschungsgemeinschaft (DFG) under
Germany’s Excellence Strategy via the Berlin Mathematics Research Center MATH+ (EXC2046/ project ID: 390685689).
\end{acknowledgments}

\bibliography{main.bib}

\widetext
\clearpage
\begin{center}
\textbf{\Large Supplemental Material: Learning Interpretable Collective Variables for Spreading Processes on Networks}
\end{center}
\setcounter{equation}{0}
\setcounter{figure}{0}
\setcounter{table}{0}
\setcounter{page}{1}
\makeatletter
\renewcommand{\theequation}{S\arabic{equation}}
\renewcommand{\thefigure}{S\arabic{figure}}

\vspace{25pt}
\textbf{\large \hspace{-15pt} S1. Supplementary Details on Binary-State Spreading Processes}
\vspace{10pt}

Our method for learning interpretable collective variables can be applied to arbitrary (binary-state) spreading processes on networks.
To illustrate this method, we apply it to two popular dynamical models: a version of the so-called \emph{voter model} and the \emph{threshold model}~\cite{Porter2016}. Both generate Markovian (i.e., memoryless) processes.

\paragraph{Voter model.}
In the \textit{continuous-time noisy voter model} (CNVM) on a simple graph containing $N$ nodes, each node $i \in \{1,\dots,N\}$ has a discrete state $x_i \in \mathbb{S}$, where $\mathbb{S}$ is a finite set of possible states.
In the context of the voter model, these states are also called \emph{opinions}.
The state space of the CNVM is $\mathbb{X} := \mathbb{S}^N$, and its elements are system states $\vec{x} = (x_1,\dots,x_N)$.
The CNVM describes the mechanism that nodes adopt the opinion of randomly chosen neighbors.
At random times occurring at a fixed rate, every node is influenced by a (uniformly chosen) random neighbor.
Hence, the rate at which a node switches to a certain opinion is proportional to the share of that opinion in its neighborhood. Simultaneously, the nodes also conduct a random opinion exploration independent of the opinions of their neighbors.
More precisely, the rate at which a node $i$ transitions from state $m \in \mathbb{S}$ to state $n \in \mathbb{S}$, $n \neq m$, is defined as
\begin{align} \label{eq:rate_votermodel}
    r_{m,n} \frac{d_{i,n}(\vec{x})}{d_{i}} + \tilde{r}_{m,n},
\end{align}
where $d_{i,n}(\vec{x})$ is the number of neighbors of node $i$ that have state $n$, $d_i$ is the degree of node $i$, and $r_{m,n}, \tilde{r}_{m,n} \geq 0$ are model parameters.
The rate $r_{m,n}$ determines how fast transitions from opinion $m$ to $n$ occur due to the above described imitation of neighbors.
The additional rate $\tilde{r}_{m,n}$ enables ``explorative'' transitions from $m$ to $n$ independently of the neighborhood and thus controls the amount of noise in the system.
Thus, by Markovianity, we have
\[
\mathrm{Prob}\left[ x_i(t+dt) = n \mid x_i(t) = m\right] = \left(r_{m,n} \frac{d_{i,n}(\vec{x})}{d_{i}} + \tilde{r}_{m,n}\right) dt + \mathcal{O}(dt^2)
\]
as~$dt \to 0$.

In all examples presented below and in the main text, we studied the CNVM with two possible opinions $\mathbb{S} = \{0, 1\}$ and model parameters
\begin{align}
    r = \begin{pmatrix}
        - & 0.99 \\
        1.01 & - \\
    \end{pmatrix}, \quad \Tilde{r} = \begin{pmatrix}
        - & 0.005 \\
        0.005 & - \\
    \end{pmatrix}.
\end{align}

\paragraph{Threshold model.}
In contrast to the voter model, where transition rates are proportional to the share of opinions in the neighborhood, the \emph{threshold model} assumes that a switch to a different opinion only occurs if that opinion is already sufficiently established in the neighborhood.
More precisely, in the threshold model a node $i$ transitions from state $m \in \{0,1\}$ to state $n = 1-m$ at the rate
\begin{equation} \label{eq:rate_threshold}
\begin{cases}
r_{m,n} + \tilde{r}_{m,n}, & \quad \frac{d_{i,n}(\vec{x})}{d_{i}} \geq b_{m,n}\\
\tilde{r}_{m,n}, & \quad \text{else}
\end{cases}
\end{equation}
where $d_{i,n}(\vec{x})$ is the number of neighbors of node $i$ that have state $n$, $d_i$ is the degree of node $i$, and $r_{m,n}, \tilde{r}_{m,n} \geq 0$ are model parameters.
The value $b_{m,n} \in [0, 1]$ is the threshold at which a node changes its opinion from $m$ to $n$.
Similarly to the voter model, the additional rates $\tilde{r}_{m,n}$ control the noise in the system.

In the examples presented below, we used the following parameters:
\begin{equation}
    r_{0,1} = r_{1,0} = 1, \quad \tilde{r}_{0,1} = \tilde{r}_{1,0} = 0.1, \quad b_{0,1} = b_{1, 0} = 0.5.
\end{equation}

\newpage
\vspace{25pt}
\textbf{\large \hspace{-15pt} S2. Supplementary Details on the Transition Manifold}
\vspace{10pt}
\paragraph{The choice of lag time.}
For a very small lag time $\tau$ compared to the timescales of the system, the transition density functions are close to Dirac distributions, i.e., $p_{\vec{x}}^\tau \approx \delta_{\vec{x}}$ for all $\vec{x}$.
Hence, the set 
\begin{equation}
    \mathbb{M}_\tau := \{p_{\vec{x}}^\tau \mid \vec{x} \in \{0,1\}^N\} \subset L^1
\end{equation}
consists of $2^N$ clearly distinct points, so that for $\vec{x} \neq \vec{y}$ we have $\lVert p_{\vec{x}}^\tau - p_{\vec{y}}^\tau \rVert \gg 0$.
For a very large time $\tau$ on the other hand, the elements $p_{\vec{x}}^\tau$ of $\mathbb{M}_\tau$ are all close to the stationary distribution $\rho \in L^1$ of the process, i.e., for all $\vec{x}$ we have $\lVert p_{\vec{x}}^\tau - \rho \rVert \approx 0$.
(Existence of the stationary distribution is guaranteed if $\tilde{r}_{0,1}, \tilde{r}_{1,0} > 0$ for the processes considered here, as this implies that the resulting Markov process is irreducible and recurrent.)
For intermediate values of $\tau$, we often observe that the set $\mathbb{M}_\tau$ is close to a $d$-dimensional submanifold $\mathbb{M} \subset L^1$ called the \emph{transition manifold} \cite{Bittracher2017, Bittracher2020}, i.e., there exists a small $\varepsilon > 0$ with
\begin{align}
    \inf_{f \in \mathbb{M}} \lVert f - p_{\vec{x}}^\tau \rVert \leq \varepsilon
\end{align}
for all $\vec{x} \in \mathbb{X}$.
Thus, the transition manifold $\mathbb{M}$ and its dimension $d$ depend on $\tau$.
However, it has been observed \cite{Bittracher2017, Bittracher2020,BiMoKoSch23} that for systems exhibiting a low-dimensional collective variable, a transition manifold of a certain dimension is robust with respect to $\tau$, in the sense that for a wide range of lag times the manifold $\mathbb{M}$ differs only slightly and its dimension remains constant.
Moreover, by Chapman--Kolmogorov, a consequence of Equation~(2) from the main document is that for $t\ge 0$ we have
\begin{equation}
    \tilde{p}^{\tau + t}_{\varphi(\vec{x})}(\vec{y}) := \int \tilde{p}^\tau_{\varphi (\vec{x})}(\vec{z})p^t_{\vec{z}}(\vec{y}) \, d\vec{z} \approx \int p^\tau_{\vec{x}}(\vec{z})p^t_{\vec{z}}(\vec{y}) \, d\vec{z} =   p^{\tau + t}_{\vec{x}}(\vec{y}),
\end{equation}
i.e., the collective variable $\varphi$ determined for the lag time $\tau$ characterizes the dynamics also for all times~$\tau +t \geq \tau$.

As many model simulations of length $\tau$ must be sampled for approximating the transition manifold in practice, it is advantageous to choose the minimum $\tau$ within the above-mentioned range of lag times.
However, if the lag time is chosen too small, the fast processes of the system will not equilibrate in that time frame, resulting in an unnecessarily large dimension $d$ and a CV that is too fine-grained.
In the case of systems with unknown time scales, it is advisable to examine trajectories to infer suitable lag times and to approximate the transition manifold for different $\tau$, comparing their dimensions $d$.
For the CNVM and the threshold model, we found that a lag time of the order such that nodes are expected to experience at least one state transition produces satisfactory results, i.e., $\tau \approx (r_{m,n} + \tilde{r}_{m,n})^{-1}$.
Nevertheless, minor modifications may be required for specific examples due to the effects of network topology.

% \paragraph{Connection to lumpability of Markov chains.}

% Choosing an adequate lag time $\tau$ results in certain states $\vec{x}$ being similar in the sense that they have almost equal transition densities $p_{\vec{x}}^\tau$, whereas for other states these densities are still quite distinct.
% In such cases all these similar states can be merged into a \textit{macro-state} that has only one associated transition density in $L^1$.
% This property is connected to the notion of \emph{lumpability} of Markov chains \cite{Bittracher2021, Simon2010, KhudaBukhsh2019}.
% Doing so for all appropriate states yields a set $\mathbb{D} \subset L^1$ of $D \ll 2^N$ densities, one for each macro-state, such that
% \begin{align}
%     \min_{f \in \mathbb{D}} \lVert f - p_{\vec{x}}^\tau \rVert \leq \varepsilon,
% \end{align}
% for all $\vec{x}$.
% The states $\vec{x}$ that are associated to the same minimizer $f \in \mathbb{D}$ belong to the same macro-state.
% We can also define these macro-states using the transition manifold theory:
% states $\vec{x}, \vec{y}$ with $\cv(\vec{x}) = \cv(\vec{y})$ have almost identical transition densities per definition of the CV $\cv$ and hence they belong to the same macro-state.
% Thus, the macro-states are given by level sets of the collective variable.

\paragraph{Approximating the transition manifold from data.}

As discussed in the main text, a CV can be obtained by learning a parametrization of the transition manifold $\mathbb{M} \subset L^1$ from data.
The approach we suggest was first presented in~\cite{Bittracher2020} and requires the calculation of distances between transition densities in the form of the maximum mean discrepancy (MMD)~\cite{Muandet2017}.
The MMD measures the difference between the means of two distributions after mapping them into a reproducing kernel Hilbert space (or \textit{feature space}) $\mathbb{H}$, which is induced by a positive definite kernel function $\kernel: \mathbb{X} \times \mathbb{X} \to \R$.
The kernel $\kernel$ also induces a \textit{feature map} $\phi: \mathbb{X} \to \mathbb{H}$ such that
\begin{align}
    \kernel(\vec{x}, \vec{y}) = \langle \phi(\vec{x}), \phi(\vec{y}) \rangle_{\mathbb{H}}.
    \label{eq:kerneltrick}
\end{align}
The MMD between two transition densities $p_{\vec{x}}^\tau$ and $p_{\vec{y}}^\tau$ is then defined as
\begin{align}
    \text{MMD}^2 \big( p_{\vec{x}}^\tau, p_{\vec{y}}^\tau \big) :=\ \big\lVert \E\big[\phi(\textbf{X}(\tau, \vec{x}))\big] - \E\big[\phi(\textbf{X}(\tau, \vec{y}))\big] \big\rVert_{\mathbb{H}}^2,
\end{align}
where $\textbf{X}(\tau, \vec{x})$ denotes a random variable with distribution~$p_{\vec{x}}^\tau$.
Here, $\E\big[\phi(\textbf{X}(\tau, \vec{x}))\big] = \int \phi(\vec{z}) p^{\tau}_{\vec{x}}(d\vec{z})$ is a Hilbert-space valued integral. Its computation is, however, not required for the evaluation of the MMD: 
Given~\eqref{eq:kerneltrick}, we can rewrite the definition of the MMD using the kernel, which is often referred to as the ``kernel trick'':
\begin{align}
    \text{MMD}^2\big( p_{\vec{x}}^\tau, p_{\vec{y}}^\tau \big) =&\ \E\Big[\kernel\big(\textbf{X}(\tau, \vec{x}), \tilde{\textbf{X}}(\tau, \vec{x})\big)\Big] + \E\Big[\kernel\big(\textbf{X}(\tau, \vec{y}), \tilde{\textbf{X}}(\tau, \vec{y})\big)\Big]\\
    &\qquad - 2\ \E\Big[\kernel\big(\textbf{X}(\tau, \vec{x}), \tilde{\textbf{X}}(\tau, \vec{y})\big)\Big], \nonumber
\end{align}
where $\textbf{X}(\tau, \vec{x}), \tilde{\textbf{X}}(\tau, \vec{x})$ are independent random variables with distribution~$p_{\vec{x}}^\tau$, and analogously for~$\vec{y}$.
This enables us to approximate the MMD by averaging over kernel function evaluations at samples of $p_{\vec{x}}^\tau$ and $p_{\vec{y}}^\tau$.

Given the set of anchor points $\vec{x}^1, \dots, \vec{x}^\numA \in \{0,1\}^N$, we define the distance matrix $\Delta \in \R^{\numA\times \numA}$, which contains the squared MMD for all pairs of states, i.e.,
\begin{align}
    \Delta_{\anchor_1, \anchor_2} := \text{MMD}^2 \big( p_{\vec{x}^{\anchor_1}}^\tau, p_{\vec{x}^{\anchor_2}}^\tau \big).
\end{align}
To estimate $\Delta$, we perform $\numS \in \N$ simulations of duration $\tau$ starting in each anchor point $\vec{x}^\anchor$ and denote the end points of these $\numS$ simulations by $\vec{y}^{(\anchor, 1)}, \dots, \vec{y}^{(\anchor, \numS)} \in \{0,1\}^N$.
Then we construct the kernel matrix $\kernelmat \in \R^{\numA\times \numA}$ with
\begin{align}
    \kernelmat_{\anchor_1, \anchor_2} := \frac{1}{\numS^2} \sum_{\sample_1, \sample_2 = 1}^\numS \kernel \big(\vec{y}^{(\anchor_1, \sample_1)}, \vec{y}^{(\anchor_2, \sample_2)} \big),
\end{align}
from which we obtain the estimation
\begin{align}
    \Delta_{\anchor_1, \anchor_2} \approx \kernelmat_{\anchor_1, \anchor_1} + \kernelmat_{\anchor_2, \anchor_2} - 2\ \kernelmat_{\anchor_1, \anchor_2}.
    \label{eq:dist_mat}
\end{align}
Finally, we apply a distance-based manifold learning algorithm to the distance matrix~$\Delta$, which yields an approximation to the $d$-dimensional CV $\cv$ evaluated at the anchor points $\vec{x}^\anchor$, i.e., $\cv(\vec{x}^1), \dots, \cv(\vec{x}^\numA) \in \R^d$.

We chose the diffusion maps method \cite{Coifman2006} as the manifold learning algorithm.
In the diffusion maps algorithm, we use a gaussian kernel with bandwidth $1$ to estimate local similarity between data points, and employ a Fokker--Planck diffusion to construct a random walk on these points. This means, $\alpha=1/2$ in~\cite{Coifman2006}. A low-dimensional representation of the data is then obtained from the dominant eigenvectors of the resulting diffusion matrix. Numerical experiments have shown that our results are robust with respect to the choice of bandwidth and diffusion type parameter~$\alpha$.

We note that the original transition manifold theory~\cite{Bittracher2017,bittracher2020weak} requires Equation~(2) from the main text to hold in a certain weighted $L^2$ norm. The justification that computations with MMD as a distance measure deliver this approximation property too has been provided in \cite[Section~3]{Bittracher2020} for dynamics exhibiting a timescale separation; see in particular Remark~3.16 therein.

\vspace{25pt}
\textbf{\large \hspace{-15pt} S3. Supplementary Details on the Algorithmic Realization}
\vspace{10pt}

In this section we provide details on each of the three steps of our method presented in the main text.

\paragraph{Step 1.}

In the first step of the method, it is crucial to choose a diverse set of dynamically relevant anchor points $\vec{x}^1, \dots, \vec{x}^\numA \in \mathbb{X}$, in the sense that their respective transition densities cover $\mathbb{M}_\tau$ sufficiently well.
For example, picking random states $\vec{x}^\anchor \sim \text{Unif}(\{0,1\}^N)$ does often not produce a desirable set of anchor points, as mostly states with about 50\% 1's are sampled (by the law of large numbers).
Thus, we would not gain any insights on the behavior of the system in states that have substantially more or less 1's than 50\%.
% (For example, on densely connected graphs all such states would have approximately the same value of the CV, as discussed in the introduction.)
Instead, we suggest to construct states $\vec{x}$ that contain communities of nodes with the same state because these are especially dynamically stable in most spreading processes. 
As configurations of nodes that involve many alternating states tend to dissolve quickly under spreading dynamics, the system is predominantly observed in states with uniform clusters (if there are clusters), and the best understanding of long term dynamics is extracted by putting emphasis on such more relevant states.
% (For other models, different types of states might be more favorable.)

In order to sample such a diverse set of anchor points $\vec{x}^1, \dots, \vec{x}^\numA$, we employ Algorithm \ref{alg:sampling}.
We start with an empty (uninitialized) state $\vec{x}$ and assign some random \emph{seed nodes} to be of state 0 and~1.
Then, we iteratively assign the state 0 to neighbors of nodes with state 0, and 1 to neighbors of nodes with state 1, until a random target count $c\in\{0,...,N\}$ of 1's and $N-c$ of 0's is reached.
For each sample $\vec{x}^\anchor$ we use a random number of seeds between $1$ and $N_{\text{seed}}= 5$. Larger or more intricate networks may require a bigger number of seeds.

{\centering
\begin{minipage}{.8\linewidth}
\begin{algorithm}[H]
    \caption{Sampling an anchor point $x$}\label{alg:sampling}
    \begin{algorithmic}[1]
    \State $x \gets \text{empty array of size $N$}$
    \State $c \gets \text{sample Unif}(\{0, \dots, N\})$ \Comment{target count of state $1$}
    \State $n \gets \text{sample Unif}(\{1, \dots, N_{\text{seed}}\})$ \Comment{number of seeds}
    \State $n \gets \min\{n, c, N-c\}$ \Comment{reduce num. of seeds (if necessary)}
    \State $i_1, \dots, i_{2n} \gets \text{random indices from $\{1, \dots, N\}$}$
    \State $x[i_1],\dots,x[i_n] \gets 0$ \Comment{initialize seed points}
    \State $x[i_{n+1}],\dots,x[i_{2n}] \gets 1$  \Comment{initialize seed points}
    
    \While{$(\text{count of state } 0) < N - c$ \textbf{or} $(\text{count of state } 1) < c$}
    \If{$(\text{count of state } 0) < N - c$}
        \State $i \gets \text{index of random uninitialized neighbor of a node with state } 0$
        \State (If no such $i$ exists, pick any random uninitialized node)
        \State $x[i] \gets 0$
    \EndIf
    
    \If{$(\text{count of state } 1) < c$}
        \State $i \gets \text{index of random uninitialized neighbor of a node with state } 1$
        \State (If no such $i$ exists, pick any random uninitialized node)
        \State $x[i] \gets 1$
    \EndIf
    \EndWhile
    \State \Return $x$
    \end{algorithmic}
\end{algorithm}
\end{minipage}
\par
}

\vspace{1cm}
\paragraph{Step 2.}
In the second step of the method, we approximate a ``parametrization'' $\cv$ of the transition manifold $\mathbb{M}$ from simulation data.
To this end, most efficient methods require the computation of local distances.
A particularly advantageous distance between densities $p^{\tau}_{\vec{x}}$ and $p^{\tau}_{\vec{y}}$ is the \emph{maximum mean discrepancy} (MMD), as it can be efficiently approximated using samples of the densities \cite{Bittracher2020, Muandet2017}.
For each anchor point $\vec{x}^\anchor$ we conduct $\numS$ simulations of the spreading process of length $\tau$, yielding $\numS$ samples for each transition density $p^{\tau}_{\vec{x}^\anchor}$.
Using this data we approximate the \textit{distance matrix} $\Delta \in \R^{\numA\times \numA}$ that contains the pairwise MMDs between the densities associated to the anchor points, which is described in detail in section S2.
Finally, we obtain evaluations of the collective variable $\cv$ at the anchor points $\vec{x}^1,\dots,\vec{x}^\numA$ by applying a manifold learning algorithm to $\Delta$. We choose the diffusion maps method \cite{Coifman2006} for this purpose.
The dimension $d$ of the CV can be obtained from the diffusion maps algorithm~\cite{CSSS08}, or can be inferred by inspecting the network structure and plots of the transition manifold.
Moreover, it is computationally cheap to test different dimensions~$d$ after the distance matrix $\Delta$ has been computed.
The output of this second step of the method is evaluations of the $d$-dimensional CV $\cv$ at the anchor points, $\cv(\vec{x}^1), \dots, \cv(\vec{x}^\numA) \in \R^d$.

\paragraph{Step 3.}
We learn a CV $\ecv$ of the form
\begin{align} \label{eq:ansatz_fun}
    \ecv(\vec{x}) = \begin{pmatrix}
\ecv_1(\vec{x})\\
\vdots \\
\ecv_d(\vec{x})
\end{pmatrix}, \quad \ecv_j(\vec{x}) = \sum_{i=1}^N \Lambda_{j, i}\; x_i,
\end{align}
by employing linear regression to find the optimal parameters $\Lambda$. To this end, we define the matrix $\cvmat := (\cv(\vec{x}^1), \dots, \cv(\vec{x}^\numA)) \in \R^{d\times \numA}$, the matrix $X \in \R^{N \times \numA}$, $X_{i,\anchor} := x_{i}^\anchor$, and note that
\begin{align}
    \ecvmat := \big(\ecv(\vec{x}^1), \dots, \ecv(\vec{x}^\numA)\big) = \Lambda X \in \R^{d\times \numA}.
\end{align}
Optimal parameters $\Lambda$, that yield a maximal correlation between our fit $\ecvmat$ and the data $\cvmat$, are then found by solving the centered linear regression problem for the rows of~$\Lambda$
\begin{align}
    \Lambda_{j,:} = \underset{\vec{\lambda} \in \R^N}{\text{argmin}} \lVert E X^T \vec{\lambda} - E \cvmat_{j,:} \rVert, \quad j=1,\dots,d. 
    \label{eq:ordinary_least_square}
\end{align}
The operator $E := I - \frac{1}{\numA} \mathbf{1} (\mathbf{1}^T) \in \R^{\numA\times \numA}$, where $I$ is the identity matrix and $ \mathbf{1} = (1, \dots, 1)^T$, centers a vector around its mean.
To prevent overfitting of the parameters $\Lambda$ to the data, we use the \emph{graph total variation} regularizer
\begin{align}
    \text{TV}(\Lambda_{j,:}) := \sum_{(i, k) \in \mathcal{E}} |\Lambda_{j, i} - \Lambda_{j, k}|,
    \label{eq:total_variation}
\end{align}
where $\mathcal{E}$ is the edge set of the graph.
This yields the generalized LASSO \cite{Tibshirani2011} linear regression problem
\begin{align}
    \Lambda_{j,:} = \underset{\vec{\lambda} \in \R^N}{\text{argmin}}\ \lVert E X^T \vec{\lambda} - E \cvmat_{j,:} \rVert + \alpha \cdot \text{TV}(\vec{\lambda}).
    \label{eq:LASSO}
\end{align}
The strength $\alpha > 0$ of the penalty can be optimized for a given data set via cross-validation \cite{Hastie2001}.
Due to the regularization, the solution $\Lambda_{j,:}$ tends to be constant within network communities, which reinforces the interpretation of $\ecv$ to measure the count of 1's in densely connected regions of the network.
The optimization problem \eqref{eq:LASSO} is convex and can be efficiently solved using off-the-shelf solvers. We employ the OSQP solver~\cite{osqp}.

For most network structures that we studied, the ansatz functions presented in \eqref{eq:ansatz_fun} resulted in CVs of good quality.
In some cases however, it can be beneficial to pre-weight each node.
For example, for the Albert--Barabási network (Example 4 in the main text) we noticed a strong correlation of the optimal (in the sense of \eqref{eq:LASSO}) weights $\Lambda$  with the node degree. This conflicts with the graph total variation regularizer because the network contains many edges between nodes of substantially different degree.
We successfully tackled this issue by incorporating a pre-weighting of each node with its degree into the ansatz functions:
\begin{align}
    \ecv(\vec{x}) = \sum_{i=1}^N \Lambda_{i}\; d_i\; x_i,
\end{align}
where $d_i$ denotes the degree of node $i$.
Transferring this pre-weighting to the regression problem \eqref{eq:LASSO} yields
\begin{align}
    \Lambda = \underset{\vec{\lambda} \in \R^N}{\text{argmin}}\ \lVert E X^T D \vec{\lambda} - E \cvmat \rVert + \alpha \cdot \text{TV}(\vec{\lambda}),
    \label{eq:LASSO_weight}
\end{align}
where $D = \text{diag}(d_1,\dots,d_N) \in \R^{N\times N}$.

If our suggested ansatz functions do not yield satisfying results for a particular network, they could also be completely replaced by functions that might be better suited for that problem.

\vspace{25pt}
\textbf{\large \hspace{-15pt} S4. Supplementary Details on the Numerical Examples}
\vspace{5pt}

In this section we provide additional details on the numerical examples presented in the main text, which all employ the continuous-time noisy voter model (CNVM) as the dynamical model.
Moreover, we apply our method to a threshold model on the same network structures as in the main text.
(For details on the models, see section S1.)
The resulting CVs are identical, which indicates that the network plays a significant role in the behavior of spreading processes on networks -- maybe even a more important role than the specific spreading model employed (voter model, threshold model, Glauber dynamics, etc.).

The \textit{Python} code for the numerical examples is available at
\begin{center}
    \url{https://github.com/lueckem/spreading-processes-CVs}.
\end{center}
The necessary model simulations were conducted using our open-source Python package \texttt{SPoNet} (Spreading Processes on Networks, \url{https://github.com/lueckem/SPoNet}).

For the calculation of the distance matrix $\Delta$, cf.~\eqref{eq:dist_mat}, we used the Gaussian kernel
\begin{align}
    \kernel (\vec{x}, \vec{y}) = \exp\Big(-\frac{\lVert \vec{x} - \vec{y} \rVert_2^2}{\sigma^2}\Big).
\end{align}
The bandwidth $\sigma$ was set to $\sqrt{N/2}$, where $N$ is the number of nodes of the respective network.
The other parameter values of the examples are summarized in Table \ref{table:params}.
The experiments were conducted on a 16-core CPU with 32 GB of memory and took less than 20 minutes each, including the simulations of the CNVM, the manifold learning, and the linear regression with cross-validation.
The most costly step ($\sim$ 70\% of runtime) in these examples was the calculation of the transition manifold parametrization, i.e., the calculation of the distance matrix \eqref{eq:dist_mat} and application of the diffusion maps algorithm.

\begin{table}
\caption{Parameter values used in the examples.}
\centering
\begin{tabular}{|c|c|c|c|c|} 
 \hline
 Parameter & stochastic block model & ring & 3-regular  & Albert--Barabási \\ [0.5ex] 
 \hline
 $N$ & 900 & 50 & 500 & 500 \\
 \hline
 $\numA$ & 2000 & 2000 & 1000 & 1000 \\ 
 \hline
 $\numS$ & 100 & 300 & 100 & 100 \\
 \hline
 $\tau$ (voter model) & 2 & 5 & 4 & 4 \\
 \hline
 $\tau$ (threshold model) & 1 & 3 & 8 & 2 \\
 \hline
 $d$ & 3 & 5 & 1 & 1 \\ [0.25ex] 
 \hline
\end{tabular}
\label{table:params}
\end{table}

\paragraph{Results for the threshold model.}
As stated before, employing the threshold model instead of the voter model results in identical transition manifolds and CVs in the three example networks.

\begin{figure}
     \centering
     \includegraphics[width=.7\linewidth]{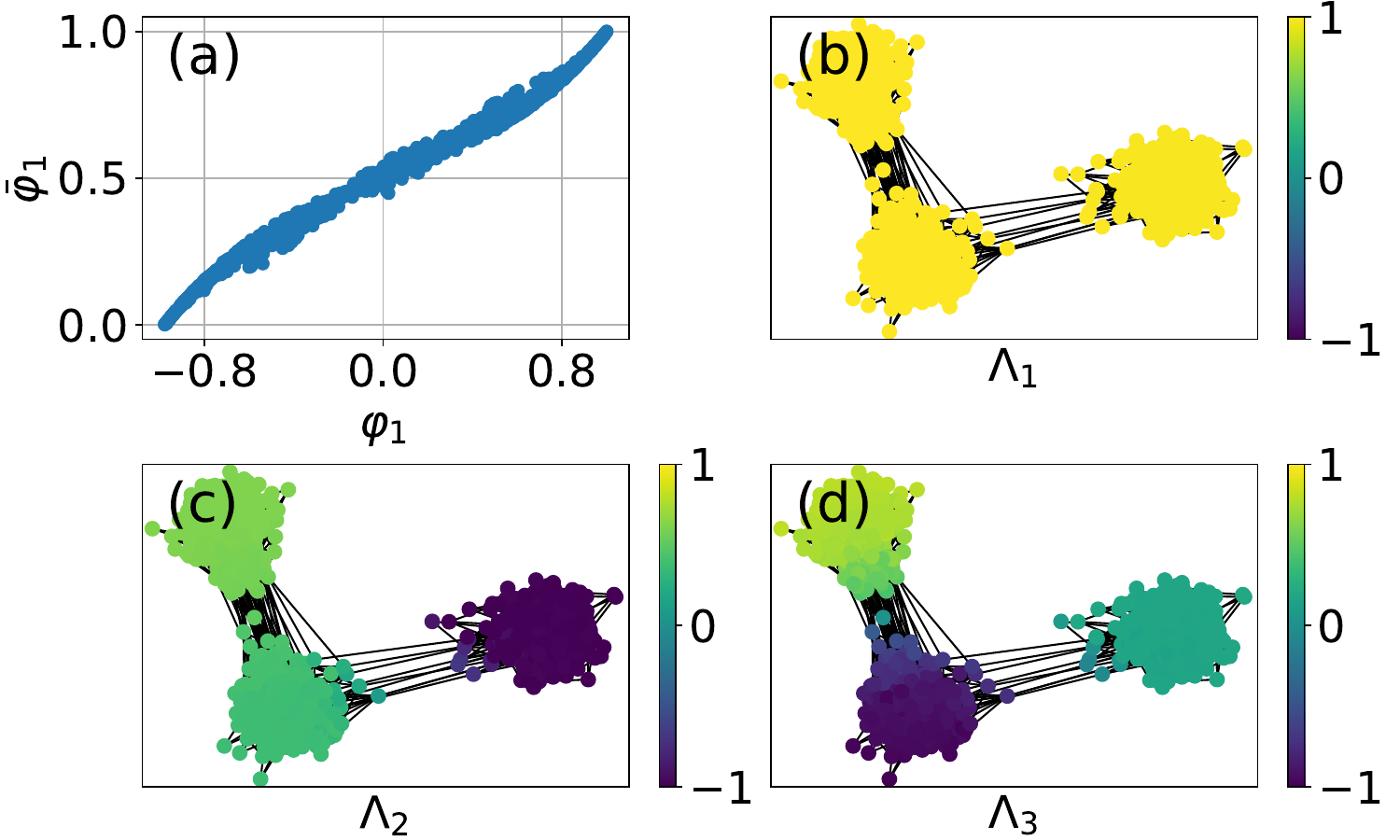}
    \caption{Optimal $\Lambda$ of the threshold model on the stochastic block model network. (a): Data $\cv_{1}(\vec{x}^k)$ versus optimal fit~$\ecv_1$. (b)-(d): Optimal $\Lambda$ entries for the respective coordinates plotted as color values on the network.}
    \label{fig:tresh_sbm_tm}
\end{figure}

For the stochastic block model network (Example 1 in the main text), the transition manifold is again a three dimensional cuboid. The CV learned by our method is illustrated in Fig.~\ref{fig:tresh_sbm_tm}.

\begin{figure}
     \centering
     \includegraphics[width=.7\linewidth]{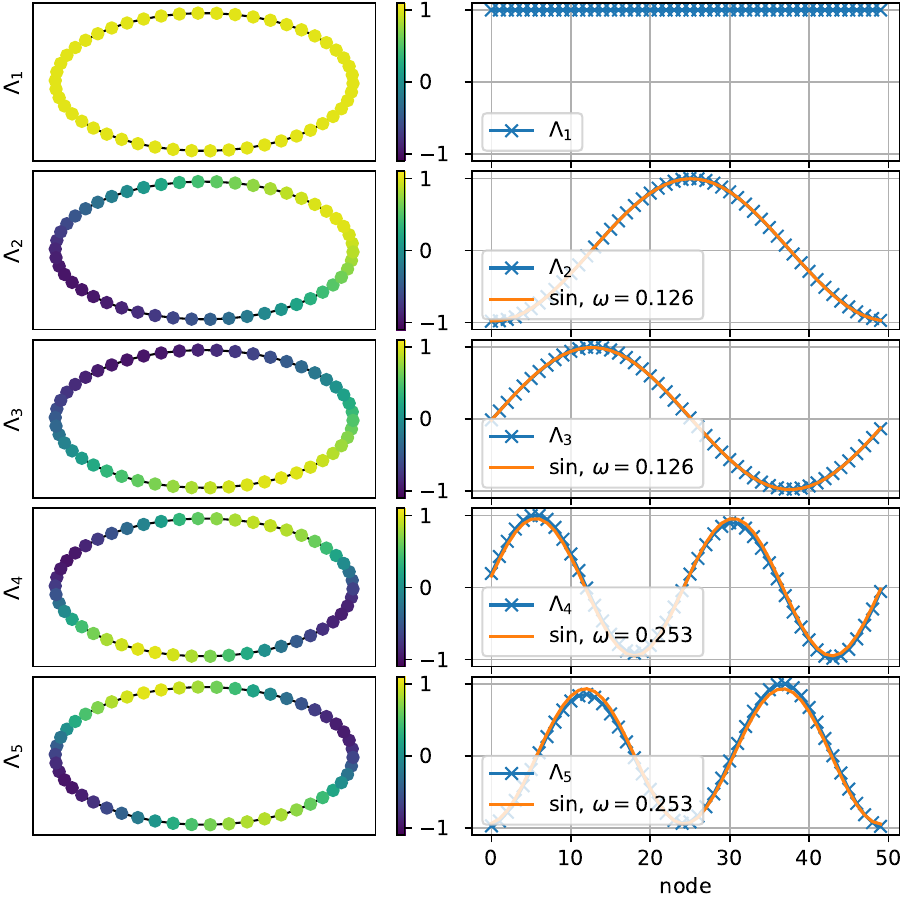}
    \caption{Left: optimal $\Lambda_{i,:}$ plotted as color values on the ring-shaped network. Right: $\Lambda_{i,:}$ (blue crosses) and a sine fit (orange line).}
    \label{fig:tresh_ring}
\end{figure}

For the ring network (Example 2 in the main text) there is no low dimensional transition manifold and the CV again represents Fourier coefficients, cf. Fig.~\ref{fig:tresh_ring}.

\begin{figure}
     \centering
     \begin{subfigure}[b]{0.48\linewidth}
         \includegraphics[width=.99\linewidth]{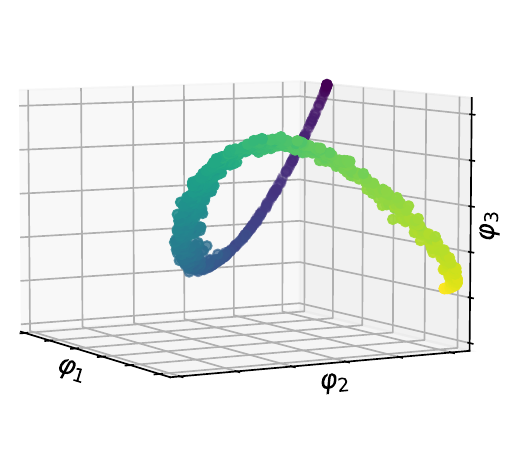}
         \caption{Voter model.}
     \end{subfigure}
     \begin{subfigure}[b]{0.48\linewidth}
         \includegraphics[width=.99\linewidth]{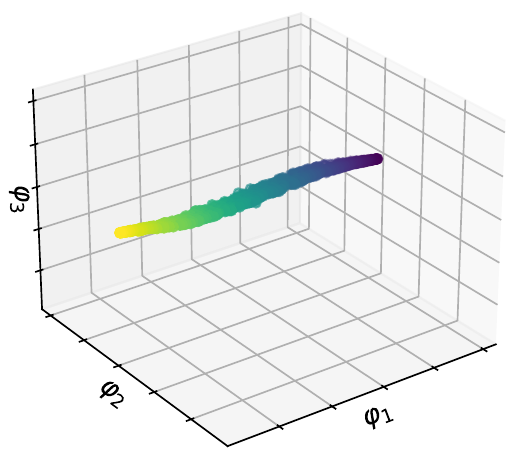}
         \caption{Threshold model.}
     \end{subfigure}
    \caption{In the Albert--Barabási example the transition manifold is one-dimensional.}
    \label{fig:ba_tm}
\end{figure}

For the Albert--Barabási network (Example 4 in the main text) the transition manifold is again one-dimensional, as shown in Fig.~\ref{fig:ba_tm}.

\paragraph{Random regular graphs.}
We have also studied the voter model and the threshold model on a uniformly random 3-regular graph with $N=500$ nodes (Example 3 in the main text). The statistical correlation between the edges (caused by regularity) poses a severe challenge to reduced modelling approaches~\cite{gleeson2011high}.

\begin{figure}
     \centering
     \begin{subfigure}[b]{0.48\linewidth}
         \includegraphics[width=.99\linewidth]{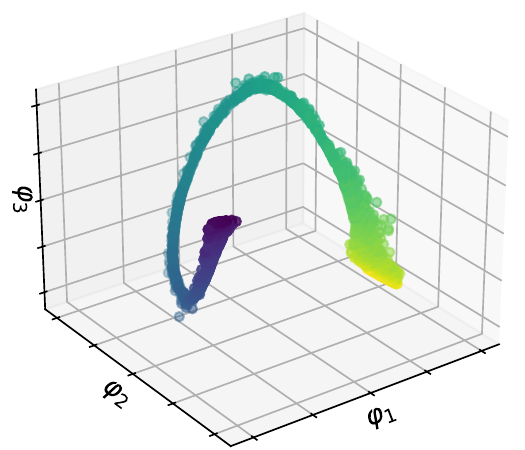}
         \caption{Voter model.}
     \end{subfigure}
     \begin{subfigure}[b]{0.48\linewidth}
         \includegraphics[width=.99\linewidth]{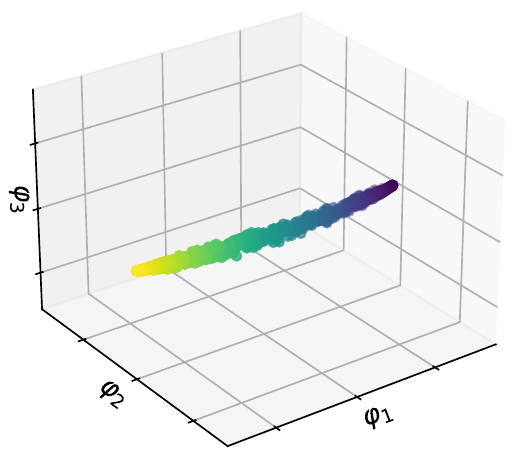}
         \caption{Threshold model.}
     \end{subfigure}
    \caption{For the random 3-regular graph the transition manifold is one-dimensional.}
    \label{fig:regular_tm}
\end{figure}

The resulting transition manifold is one-dimensional and the learned CV calculates the global share of $1$'s, cf. Fig~\ref{fig:regular_tm}.
Hence, the CV is identical to the optimal CV in the complete graph and Erd\H{o}s--Rényi random graph (with sufficiently large edge density) setting~\cite{Luecke2022}.
However, the dynamics on random 3-regular graphs behaves (quantitatively) differently than on a complete graph. In particular, the known mean-field equation, which constitutes the large population limit of the voter model on complete networks, is not valid for random regular graphs.
We are not aware of a similar dynamical equation for the share of 1's in the random regular graph setting but the one-dimensionality of the transition manifold indicates its existence.

\vspace{25pt}
\textbf{\large \hspace{-15pt} S5. Validation}
\vspace{5pt}

Recall that the transition manifold approach is seeking for a low-dimensional parametrization $\cv$ of the set of all transition densities $p_{\vec{x}}^\tau$, $\vec{x} \in \{0,1\}^N$. Thus, for two states $\vec{x}$ and $\vec{y}$ the distance between $\smash{p_{\vec{x}}^\tau}$ and $\smash{p_{\vec{y}}^\tau}$ should correlate with the distance between $\cv(\vec{x})$ and~$\cv(\vec{y})$.
Thus, the quality of a collective variable $\cv$ can be assessed using the following heuristic.
Given a small $\varepsilon > 0$ and two states $\vec{x}^1, \vec{x}^2 \in \{0,1\}^N$ with $\cv(\vec{x}^1) \approx \cv(\vec{x}^2)$, we define the time $t^*$ as 
\begin{align}
    t^* := \inf\{t \geq 0 \mid \forall \tau \geq t: \lVert p_{\vec{x}^1}^\tau - p_{\vec{x}^2}^\tau \rVert < \varepsilon\}.
\end{align}
This is well-defined because we assume that all $p_{\vec{x}}^\tau$ converge to a unique stationary distribution for $\tau \to \infty$.
If $t^*$ is rather large, this implies that the CV is very coarse because initial states with similar CV values may lead to quite different behavior over long timescales.
But a good CV has the property that starting the system in $\vec{x}^1$ versus in $\vec{x}^2$ should make almost no difference after a short time, and hence, good CVs exhibit a small $t^*$.
However, this property alone does not sufficiently characterize the quality of the CV (for example, $\cv(\vec{x}) = \vec{x}$ implies $t^* = 0$, but it does not simplify the state at all).
Hence, we additionally define a third state $\vec{x}^3$ with a clearly distinct reduced state $\lVert \cv(\vec{x}^3) - \cv(\vec{x}^1) \rVert \gg 0$. Then the CV is of good quality if the associated distributions $p_{\vec{x}^1}^\tau$ and $p_{\vec{x}^3}^\tau$ are also substantially different for a long time, i.e.,
\begin{align}
    \hat{t} := \inf\{t \geq 0 \mid \forall \tau \geq t: \lVert p_{\vec{x}^1}^\tau - p_{\vec{x}^3}^\tau \rVert < \varepsilon\}
\end{align}
should be large.
If $\hat{t}$ is rather small, this implies that the CV is too fine-grained because it assigns different values to initial states that lead to a similar dynamics after a short time.
The best possible CV achieves both the smallest $t^*$ for all choices of $\vec{x}^1, \vec{x}^2$ and the largest $\hat{t}$ for all choices of $\vec{x}^1, \vec{x}^3$, as it exactly filters out the state information with a short timescale impact, but keeps the information that leads to fundamentally different dynamics on a long timescale.

We propose the following numerical method for validation. After the approximation $\ecv$ of the collective variable has been calculated by our method, we pick three states $\vec{x}^1, \vec{x}^2, \vec{x}^3 \in \{0,1\}^N$ as discussed above, i.e, such that $\ecv(\vec{x}^1) \approx \ecv(\vec{x}^2)$ and $\ecv(\vec{x}^3)$ is substantially different.\footnote{Given $\vec{x}^1$, we sample $\vec{x}^2$ using a Markov chain Monte Carlo method. Starting with a uniformly random $\vec{x}^2$, we randomly flip states of nodes until $\ecv(\vec{x}^1) \approx \ecv(\vec{x}^2)$.}
Then we compare the distributions $p_{\vec{x}^1}^\tau$, $p_{\vec{x}^2}^\tau$, and $p_{\vec{x}^3}^\tau$ via their maximum mean discrepancy (MMD)
\begin{align}
    \text{MMD}^2(\vec{x}^i, \vec{x}^j; t) :=&\ \E\Big[\kernel\Big(\textbf{X}(t, \vec{x}^i), \tilde{\textbf{X}}(t, \vec{x}^i)\Big)\Big]
    + \E\Big[\kernel\Big(\textbf{X}(t, \vec{x}^j), \tilde{\textbf{X}}(t, \vec{x}^j)\Big)\Big] \nonumber\\
    &\qquad - 2\ \E\Big[\kernel\Big(\textbf{X}(t, \vec{x}^i), \tilde{\textbf{X}}(t, \vec{x}^j)\Big)\Big],
    \label{eq:mmd_validate_2}
\end{align}
where $\textbf{X}(t, \vec{x})$ and $\tilde{\textbf{X}}(t, \vec{x})$ are independent random variables with distribution $p_{\vec{x}}^t$, and $\kernel(\cdot , \cdot)$ is a Gaussian kernel.
If the learned CV $\ecv$ is good, we expect that $\text{MMD}(\vec{x}^1, \vec{x}^2; t)$ goes to $0$ quickly (as $t^*$ is small), while both $\text{MMD}(\vec{x}^1, \vec{x}^3; t)$ and $\text{MMD}(\vec{x}^2, \vec{x}^3; t)$ stay large for a long time (as $\hat{t}$ is large).
We estimate the three MMDs by replacing the expectation in \eqref{eq:mmd_validate_2} with averages from model simulations.
To prove the quality of the learned CV $\ecv$, one would have to conduct this experiment for all possible choices of $\vec{x}^1, \vec{x}^2, \vec{x}^3$, which is not feasible.
However, we argue that doing it for a few (random) choices is a good indicator of whether the CV is correct.

A weaker (but easier to compute) indicator of the quality of a CV is given by the differences between the projected distributions, i.e.,
\begin{align}
    \text{MMD}^2_\cv(\vec{x}^i, \vec{x}^j; t) :=&\ \E\Big[\kernel\Big(\ecv(\textbf{X}(t, \vec{x}^i)), \ecv(\tilde{\textbf{X}}(t, \vec{x}^i))\Big)\Big]
    + \E\Big[\kernel\Big(\ecv(\textbf{X}(t, \vec{x}^j)), \ecv(\tilde{\textbf{X}}(t, \vec{x}^j))\Big)\Big] \nonumber\\
    &\qquad - 2\ \E\Big[\kernel\Big(\ecv(\textbf{X}(t, \vec{x}^i)), \ecv(\tilde{\textbf{X}}(t, \vec{x}^j))\Big)\Big].
    \label{eq:mmd_validate}
\end{align}
(The kernels $\kappa$ in \eqref{eq:mmd_validate_2} and in~\eqref{eq:mmd_validate} are clearly different objects, as the state $\vec{x}$ and reduced state $\ecv(\vec{x})$ have different dimensions.)
In contrast to \eqref{eq:mmd_validate_2}, we expect that $\text{MMD}_\cv(\vec{x}^1, \vec{x}^2; t)$ is close to $0$ for all times $t$, as states with the same CV value should lead to identical effective (projected) dynamics.
The differences $\text{MMD}_\cv(\vec{x}^1, \vec{x}^3; t)$ and $\text{MMD}_\cv(\vec{x}^2, \vec{x}^3; t)$ should again be large for small and intermediate $t$. For very large $t$, even they should approach zero as the distribution of all $\textbf{X}(t, \vec{x}^j)$ converge to the stationary distribution.

\begin{figure}
     \centering
     % \vspace{-1cm}
     \begin{subfigure}[b]{0.22\linewidth}
         \includegraphics[width=.99\linewidth]{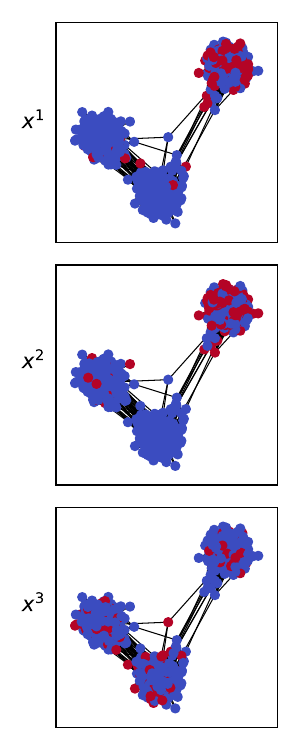}
         \caption{States $\vec{x}^1, \vec{x}^2, \vec{x}^3$.}
     \end{subfigure}
     \begin{subfigure}[b]{0.31\linewidth}
         \includegraphics[width=.99\linewidth]{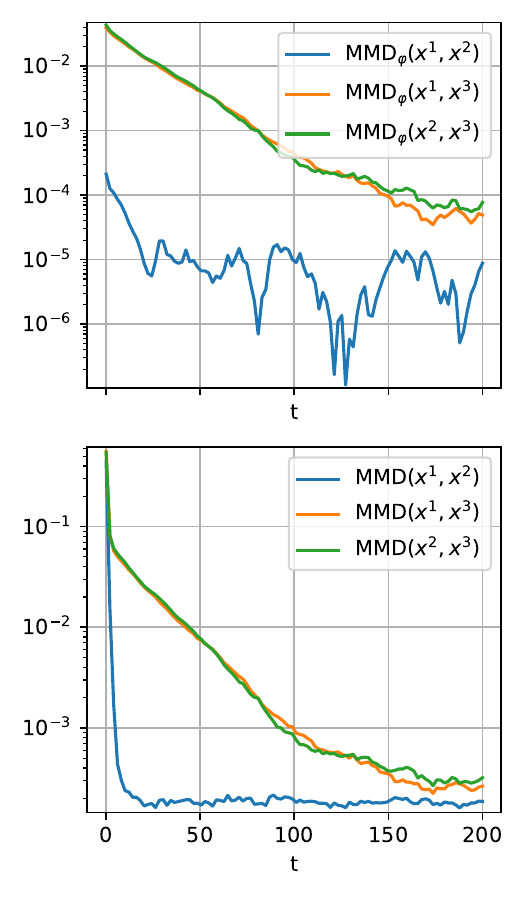}
         \caption{Voter model.}
     \end{subfigure}
     \begin{subfigure}[b]{0.31\linewidth}
         \includegraphics[width=.99\linewidth]{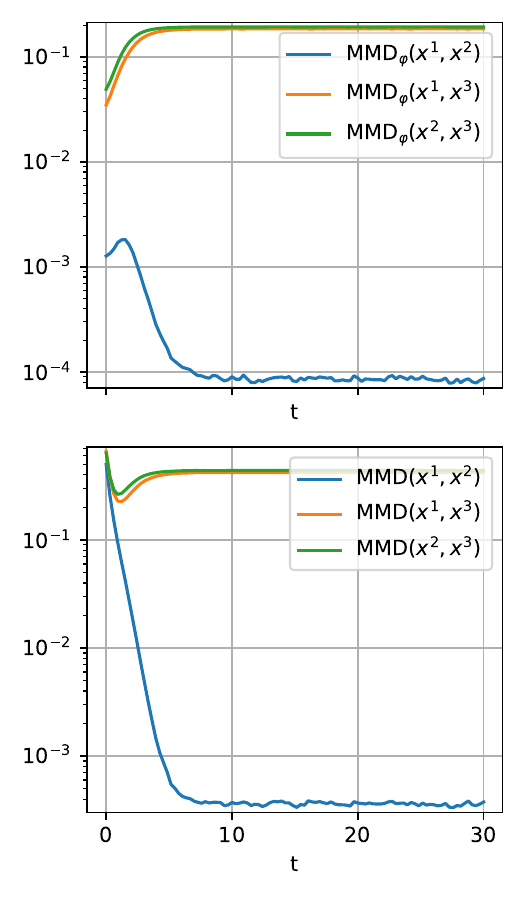}
         \caption{Threshold model.}
     \end{subfigure}
    \caption{Validation for the stochastic block model example. For a definition of $\text{MMD} $ and $\text{MMD}_\cv$ see~\eqref{eq:mmd_validate_2} and \eqref{eq:mmd_validate}.}
    \label{fig:validate_sbm}
\end{figure}

\paragraph{Results.}
For the stochastic block model example, we picked states $\vec{x}^1$ and $\vec{x}^2$ such that the share of 1's is 10\% in cluster 1, 0\% in cluster 2, and 50\% in cluster 3, see Fig.~\ref{fig:validate_sbm}. In state $\vec{x}^3$ the total number of 1's is the same, but they are distributed uniformly, irrespective of the clusters.
For both the voter model dynamics and the threshold model dynamics the time $t^*$ is very small ($t^* \approx 5)$ compared to $\hat{t}$.
This indicates that the learned CV is good, as discussed above.
An interesting difference between the voter model and the threshold model on this network is that the metastable states (where each cluster is dominated by one opinion) persist much longer in the threshold model.
As $\vec{x}^1$ and $\vec{x}^3$ typically move into different metastable states, their MMD stays large much longer in the threshold model.

\begin{figure}
     \centering
     \vspace{-0.5cm}
     \begin{subfigure}[b]{0.22\linewidth}
         \includegraphics[width=.99\linewidth]{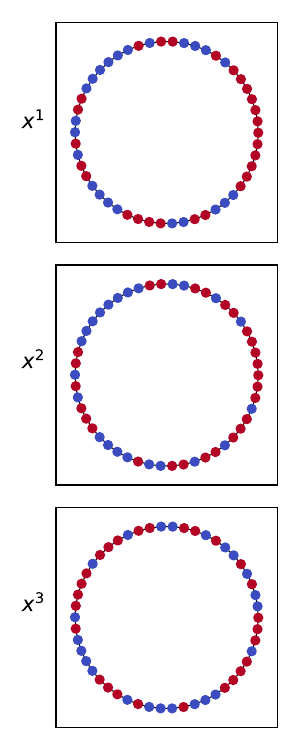}
         \caption{States $\vec{x}^1, \vec{x}^2, \vec{x}^3$.}
     \end{subfigure}
     \begin{subfigure}[b]{0.31\linewidth}
         \includegraphics[width=.99\linewidth]{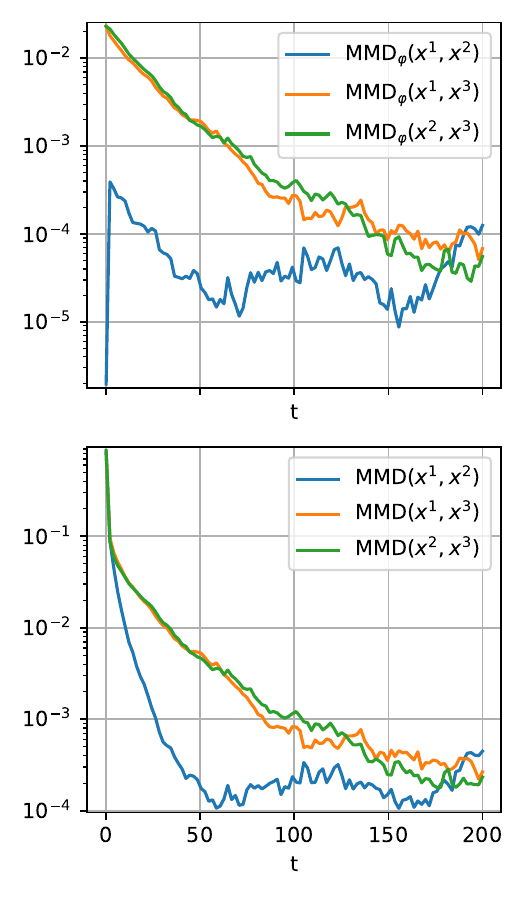}
         \caption{Voter model.}
     \end{subfigure}
     \begin{subfigure}[b]{0.31\linewidth}
         \includegraphics[width=.99\linewidth]{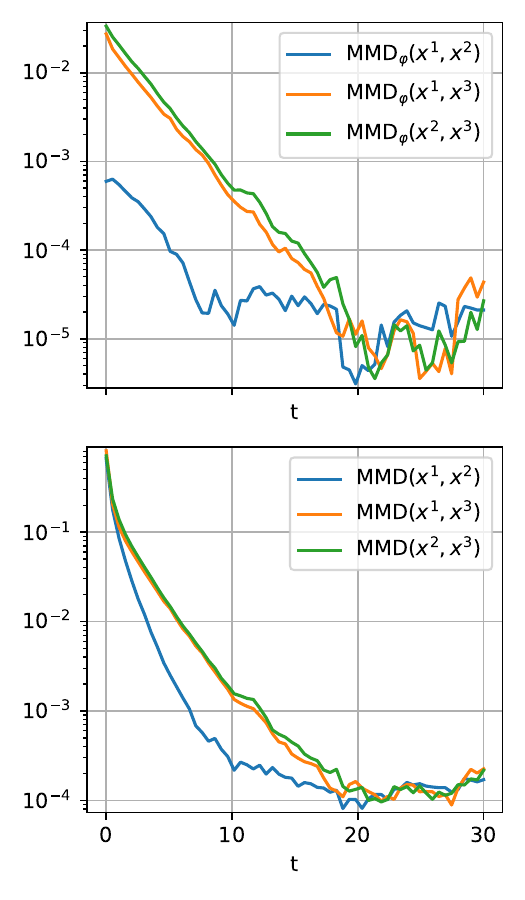}
          \caption{Threshold model.}
     \end{subfigure}
    \caption{Validation for the ring network example. For a definition of $\text{MMD} $ and $\text{MMD}_\cv$ see~\eqref{eq:mmd_validate_2} and \eqref{eq:mmd_validate}.}
    \label{fig:validate_ring}
\end{figure}

For the ring network, state $\vec{x}^1$ is chosen randomly (i.i.d.\ for every node), $\vec{x}^2$ is chosen such that the collective variable matches, and $\vec{x}^3$ is given by a random permutation of node states of $\vec{x}^1$, see Fig.~\ref{fig:validate_ring}. Hence, state $\vec{x}^1$ and $\vec{x}^3$ have the same number of~1's.
Here, the time $t^*$ is rather large compared to $\hat{t}$, which shows that this CV is too coarse.
This is not surprising because we capped the CV dimension at five, whereas the transition manifold approach suggested a higher dimension, see the main text.
One could improve the quality of this CV (i.e., reduce $t^*$) at the cost of increasing its dimension and hence complexity.

\begin{figure}
     \centering
     % \vspace{-1cm}
     \begin{subfigure}[b]{0.22\linewidth}
         \centering
         \includegraphics[width=.99\linewidth]{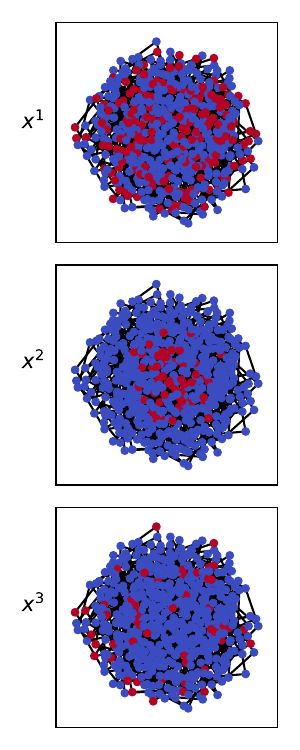}
         \caption{States $\vec{x}^1, \vec{x}^2, \vec{x}^3$.}
     \end{subfigure}
     \begin{subfigure}[b]{0.31\linewidth}
         \centering
         \includegraphics[width=.99\linewidth]{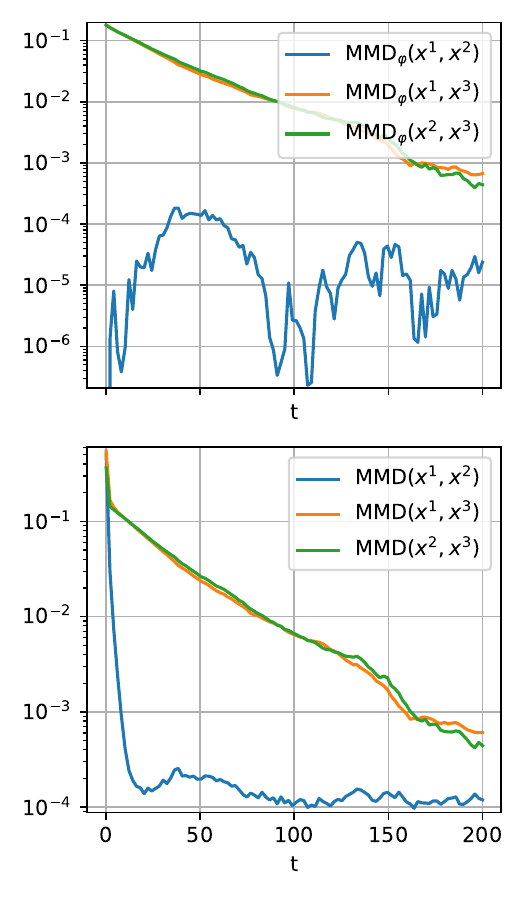}
         \caption{Voter model.}
     \end{subfigure}
     \begin{subfigure}[b]{0.31\linewidth}
         \centering
         \includegraphics[width=.99\linewidth]{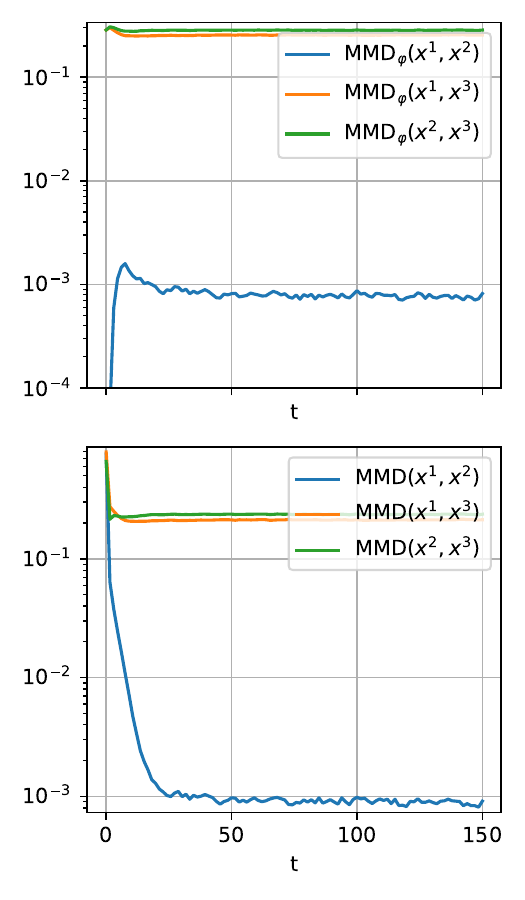}
         \caption{Threshold model.}
     \end{subfigure}
    \caption{Validation for the Albert--Barabási network example. For a definition of $\text{MMD} $ and $\text{MMD}_\cv$ see~\eqref{eq:mmd_validate_2} and \eqref{eq:mmd_validate}.}
    \label{fig:validate_ba}
\end{figure}

For the Albert--Barabási network, state $\vec{x}^2$ is such that the top 10\% of nodes with largest degree are in state 1, whereas in $\vec{x}^3$ the 10\% of nodes with smallest degree are in state 1, see Fig.~\ref{fig:validate_ba}.
Hence, the total number of 1's is identical, but the CV, which measures the degree-weighted count of 1's, differs substantially.
State $\vec{x}^1$ is chosen to have the same degree-weighted count of 1's as state~$\vec{x}^2$.
This experiment indicates that the learned CV is good because $t^*$ is small compared to $\hat{t}$.
Moreover, we again observe that due to the longer persistence of metastable states in the threshold model, the MMD between $\vec{x}^1$ and $\vec{x}^3$ stays much larger in the threshold model than in the voter model.

\begin{figure}
     \centering
     \vspace{-0.5cm}
     \begin{subfigure}[b]{0.22\linewidth}
         \centering
         \includegraphics[width=.99\linewidth]{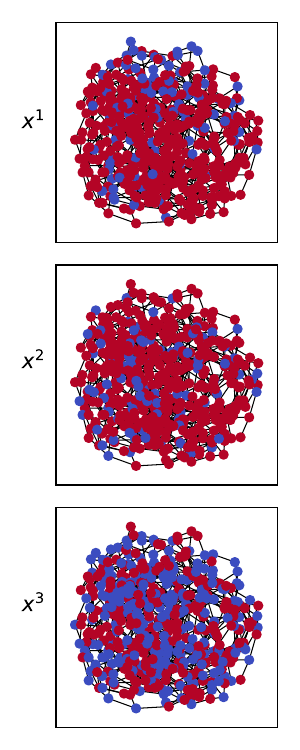}
         \caption{States $\vec{x}^1, \vec{x}^2, \vec{x}^3$.}
     \end{subfigure}
     \begin{subfigure}[b]{0.31\linewidth}
         \centering
         \includegraphics[width=.99\linewidth]{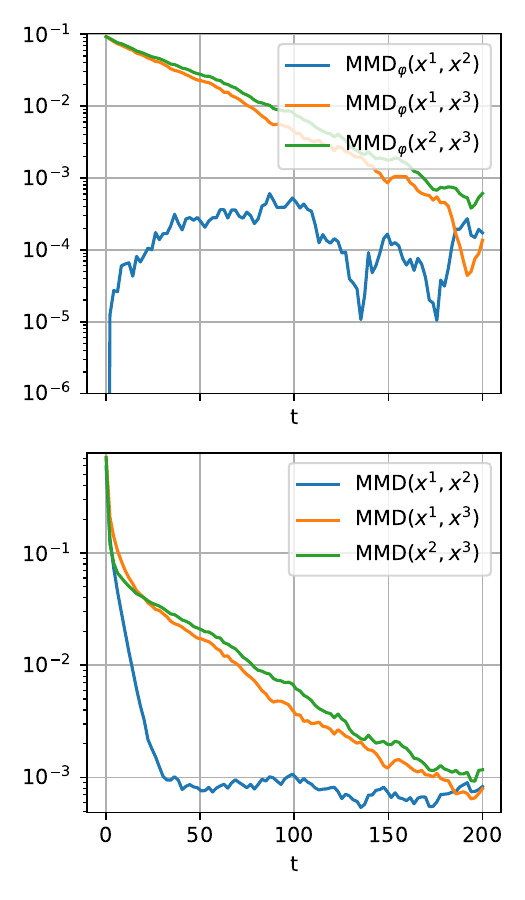}
         \caption{Voter model.}
     \end{subfigure}
     \begin{subfigure}[b]{0.31\linewidth}
         \centering
         \includegraphics[width=.99\linewidth]{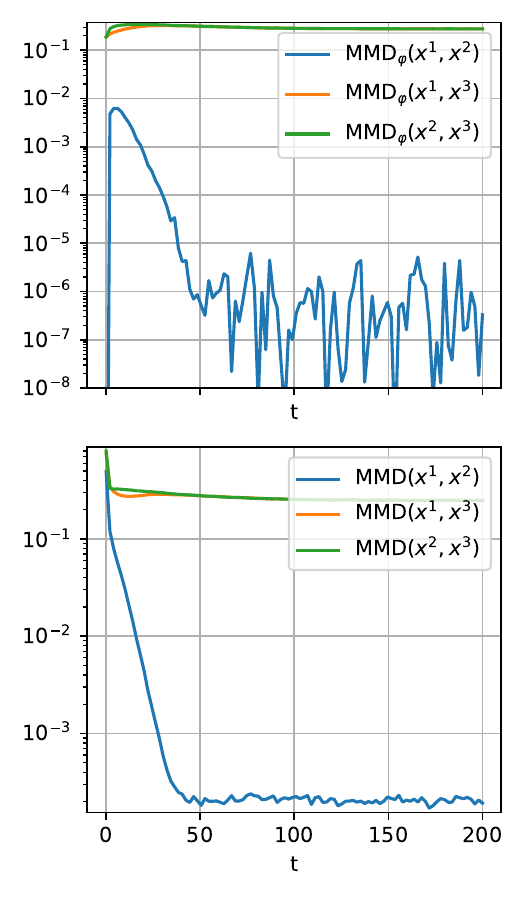}
         \caption{Threshold model.}
     \end{subfigure}
    \caption{Validation for the random 3-regular network. For a definition of $\text{MMD} $ and $\text{MMD}_\cv$ see~\eqref{eq:mmd_validate_2} and \eqref{eq:mmd_validate}.}
    \label{fig:validate_regular}
\end{figure}

Finally, we validate that the simple share of 1's is indeed a good CV in the case of the random 3-regular network in Fig.~\ref{fig:validate_regular}.
States $\vec{x}^1$ and $\vec{x}^2$ are random permutations with the same number of 1's, whereas $\vec{x}^3$ has a different number of 1's.
The time $t^*$ is again small compared to $\hat{t}$, which implies that the CV is good.

\paragraph{Robustness of the results.}
For all considered network topologies and spreading processes on them we have tested the robustness of our results with respect to variations in the model parameters~$r,\tilde{r}$. We observed that the results reported here all remained valid qualitatively as long as the parameter values were not varied by many orders of magnitude.

% \bibliographystyle{abbrv}
% \bibliography{main.bib}

\end{document}